\documentclass[twocolumn]{aastex63}

\usepackage{graphicx}
\usepackage{color, colortbl}
\usepackage{float}
\usepackage{courier}

\usepackage{tikz}

\received{2023 September 21}
\revised{2024 February 20}
\accepted{2024 February 29}

\shorttitle{Variability in NGC 4388}
\shortauthors{Gediman et al.}

\begin{document}

\title{Test for Echo: X-ray Reflection Variability in the Seyfert-2 AGN NGC 4388}

\author{Ben Gediman}
\email{bgediman@umich.edu}
\affiliation{Department of Astronomy, University of Michigan, 1085 South University Avenue, Ann Arbor, MI 48109, USA}

\author{Jon M. Miller}
\affiliation{Department of Astronomy, University of Michigan, 1085 South University Avenue, Ann Arbor, MI 48109, USA}

\author{Abderahmen Zoghbi}
\affiliation{Department of Astronomy, University of Maryland, College Park, MD, 20742, USA}
\affiliation{HEASARC, Code 6601, NASA/GSFC, Greenbelt, MD, 20771, USA}
\affiliation{CRESST II, NASA Goddard Space Flight Center, Greenbelt, MD, 20771, USA}

\author{Paul Draghis}
\affiliation{Department of Astronomy, University of Michigan, 1085 South University Avenue, Ann Arbor, MI 48109, USA}

\author{Zaven Arzoumanian}
\affiliation{NASA Goddard Space Flight Center, Code 662, Greenbelt, MD 20771, USA}

\author{W. N. Brandt}
\affiliation{Department of Astronomy and Astrophysics, 525 Davey Lab, The Pennsylvania State University, University Park, PA 16802, USA}
\affiliation{Institute for Gravitation and the Cosmos, The Pennsylvania State University, University Park, PA 16802, USA}
\affiliation{Department of Physics, 104 Davey Lab, The Pennsylvania State University, University Park, PA 16802, USA}

\author{Keith Gendreau}
\affiliation{NASA Goddard Space Flight Center, Code 662, Greenbelt, MD 20771, USA}

\begin{abstract}
We report on a study of the narrow Fe~K$\alpha$ line and reflection spectrum in the well-known Seyfert-2 AGN, NGC 4388.  X-ray spectra summed from two extensive NICER monitoring campaigns, separated by years, show strong evidence of variation in the direct continuum and reflected emission, but only small variations in the obscuring gas.  Fits to the spectra from individual NICER observations find a strong, positive correlation between the power-law photon index, $\Gamma$, and direct flux that is commonly observed in unobscured AGN.  A search for a reverberation lag between the direct and reflected spectra---dominated by the narrow Fe~K$\alpha$ emission line---measures a time scale of $t = 16.37^{+0.46}_{-0.38}$~days, or a characteristic radius of $r=1.374_{-0.032}^{+0.039}\times10^{-2}$~pc $=3.4_{-0.1}^{+0.1}\times10^4\;GM/c^2$. Only one cycle of this tentative lag is observed, but it is driven by a particularly sharp drop in the direct continuum that leads to the subsequent disappearance of the otherwise prominent Fe~K$\alpha$ line.  Physically motivated fits to high-resolution Chandra spectra of NGC 4388 measure a line production radius of $r =2.9^{+1.2}_{-0.7}~\times 10^{4}~GM/c^{2}$, formally consistent with the tentative lag.   The line profile also prefers a Compton-thick reflector, indicating an origin in the disk and/or thick clumps within a wind.  We discuss the strengths and weaknesses of our analysis and methods for testing our results in future observations, and we note the potential for X-ray reverberation lags to constrain black hole masses in obscured Seyferts wherein the optical broad line region is not visible.
\end{abstract}

\keywords{black holes, active galactic nuclei, X-ray active galactic nuclei}

\section{Introduction \label{sec:intro}}
NGC 4388 is a heavily obscured Seyfert-2 galaxy. At a distance of $d = 18.0$~Mpc (\citealt{sorce2014}), it is one of the closest examples of this kind of active galactic nucleus (AGN), and one of the brightest in X-rays.  Non-thermal X-ray emission from the nucleus has been detected out to nearly 500~keV (\citealt{beckmann2004}).  The obscuration that hides the central engine is not Compton-thick, but is only a factor of a few below this key threshold (see, e.g., \citealt{miller2019}).  Although the mass of the black hole cannot be determined using optical reverberation techniques (see, e.g., \citealt{peterson2004}), ${\rm H}_{2}{\rm O}$ megamaser emission in its disk has enabled a precise mass measurement: $M_{BH} = 8.4\,\pm\,0.2\times 10^{6}~M_{\odot}$ (\citealt{kuo2011}).  

This mass measurement and the high X-ray flux of NGC 4388 present some unique opportunities.  Flux variability time scales in the direct and reprocessed emission from NGC 4388 can be translated to gravitational radii ($r_{g} = GM/c^{2}$) with unusual precision, and can therefore help to resolve the nature and origin of the obscuring ``torus'' geometry (e.g., \citealt{antonucci1993}).  At least one past study has found that the obscuration in NGC 4388 can vary by orders of magnitude on a time scale of just hours (\citealt{elvis2004}), suggesting that it extends much closer to the black hole than the parsec-scale geometry that is classically envisioned (also see \citealt{risaliti2002}).

The narrow Fe~K$\alpha$ emission line is the strongest atomic feature in the X-ray spectra of Seyfert AGN (for a recent review, see \citealt{gallo2023}). Owing to its large geometric covering factor of the torus, at least some Fe~K$\alpha$ line flux must originate {\it on its face}, no matter its distance from the central engine.  However, recent studies of some Seyfert-1 AGN suggest that the bulk of the line flux is instead produced within the optical broad line region (BLR), or at even smaller radii (see, e.g., \citealt{miller2018}, \citealt{zoghbi2019}, \citealt{noda2023}).  Those studies are particularly sensitive to the {\it innermost} extent of the torus. The high flux observed in NGC 4388 has even permitted a tentative detection of Ni~K$\alpha$ emission (\citealt{fukazawa2016}).  However, sensitive observations of NGC 4388 have not revealed relativistic reflection from the innermost disk (\citealt{kamraj2017}, \citealt{yaqoob2023}).  

It is possible that the line of sight determines which geometry contributes the bulk of the Fe K line flux, and NGC 4388 may provide a key angle on this question.  The observed line flux may be dominated by a distant part of the torus along our line of sight, since the BLR is obscured.  Although the innermost extent of the torus may only be a few times larger than the optical BLR (e.g., \citealt{minezaki2019}), its radial extent may span orders of magnitude, out to the scale of parsecs (or larger).  Complete quenching of the line flux from such a geometry should be nearly impossible in this circumstance, and variations should be slow and muted relative to the central engine and the response of the BLR.  If the innermost edge of the obscuring gas instead extends as close to the black hole as prior studies may indicate, and if the innermost edge contributes the bulk of the narrow Fe~K line flux, it may also vary on short time scales. We note that weak Fe~K emission on kpc scales has been detected in NGC 4388 (\citealt{iwasawa2003}, \citealt{yi2021}), but this emission is not expected to vary on human time scales.

The spectral resolution and collecting area of NICER (\citealt{gendreau2016}) enable spectroscopic monitoring of bright Seyferts, opening new windows on the accretion flow geometry in these AGN.  Motivated by this opportunity and the strong variations previously found in NGC 4388, we obtained a series of short monitoring exposures to search for variations in the obscuring column and Fe~K$\alpha$ line flux as diagnostics of the torus.  We also summed these new observations to create a very sensitive time-averaged spectrum, similar to another series of prior observations (\citealt{miller2019}). 

In this paper, we utilize both the prior and new sets of NICER observations to leverage the variability of spectral features to better understand the inner accretion flow in NGC 4388. We additionally use archival Chandra observations to compare these results with those derived from the detailed Fe K line profile in NGC 4388. In Section \ref{sec:obs}, we detail the sets of observations and our reduction of the data.  Our analysis and results are presented in Section \ref{sec:analysis}.  Finally, we discuss the strengths and limitations of our work in Section \ref{sec:discussion}, and speculate on the potential of future observations.

\section{Observations and Data Reduction \label{sec:obs}}
\subsection{NICER}
\begin{deluxetable*}{c|cccc}
    \tablecaption{Summary of Observations}
    \label{table:observation_log}
    \tablewidth{\textwidth} 
    \tablehead{
        \colhead{Mission} & \colhead{Observations} & \colhead{ObsIDs} & \colhead{Dates} & \colhead{Net Exposure Time (ks)}
    }
    \startdata
    NICER & 34 & 1117010101--1117010134 & 2017-12-03 -- 2018-01-11 & 103.3 \\
    & 11 & 5117010101--5117010111 & 2022-06-09 -- 2022-06-19 & 96.0 \\
    Chandra & 2 & 9276--9277 & 2008-04-16 -- 2008-04-24 & 269 \\
    \enddata
    \tablecomments{Details of the observations used in this analysis. Our main spectral analysis utilizes two epochs of NICER observations, separated by roughly 4.5 years, with each amounting to about 100 ks of total exposure time. We additionally used two Chandra observations, totalling 269 ks, for supplementary analysis.}
\end{deluxetable*}

NICER observations of NGC 4388 can be grouped into three sets. The first set was obtained near the start of the NICER mission, beginning on 2017 December 03 (MJD 58090) and continuing almost daily until 2018 January 11 (MJD 58129). A series of 34 observations accumulated 109.8~ks of total exposure time. The second set consists of 10 observations from 2018 December 05 (MJD 58457) to 2019 April 04 (MJD 58577) for a total of 9.2~ks. An analysis of the time-averaged spectrum of these two campaigns is reported in \cite{miller2019}. The third set includes 11 daily observations from 2022 June 09 (MJD 59739) to 2022 June 19 (MJD 59739), for a total of 100.0~ks.

In this investigation, we were concerned with separately exploring variability on short and long time scales, on the order of days and years, respectively. The earliest and latest of these three epochs suited these goals, since both have roughly 100~ks of total exposure time and they are separated from each other by over four years. Because the observations in the middle epoch are spaced relatively far apart and have comparatively low exposure time, we chose not to include them in our analysis. While there is some variation, the earliest epoch had a typical exposure time of about 1.5--4 ks per observation. Compared to the latest epoch, with the majority of observations having exposure times greater than 9 ks, the early epoch covers a longer time period and is therefore additionally well suited to investigating variability on the time scale of days.

Each of the 45 observations from the first and third epochs (ObsID 1117010101--34 and 5117010101--11) was processed using \texttt{HEASoft 6.29c}. The data reduction was accomplished using \texttt{NICERDAS 8c}, with the ``xti20210707'' \texttt{CALDB} calibration file. The more recent observations all had a considerably larger high-energy background that was not flagged by the standard \texttt{nicerl2} pipeline. Using the program \texttt{nicermaketime}, we enforced a cap on the magnetic cutoff rigidity (\texttt{COR\_SAX}), which is an index describing the level of particle background, as well as the number of ``overshoot'' events that are caused by high energy photons, charged particles, and cosmic rays. In particular, we imposed the \texttt{nicermaketime} parameters \texttt{cor\_range=2.0-*} (no default value) and \texttt{overonly\_range=0-1} (default 0--1.5) on all observations.  We additionally excised two isolated time intervals ($\sim$100 s each) that were responsible for low-energy background flares in ObsIDs 1117010133 and 1117010134 that were not caught by the previous methods. 

After this screening, the net exposure time was 103.3 ks for the first epoch and 96.0 ks for the third epoch. The individual observations from each period were also combined using \texttt{nicermergeclean}, and then all 47 spectra (45 individual observations plus 2 time averages) were extracted using \texttt{XSELECT v2.4m}. After spectra had been extracted, ARF and RMF response files were generated using \texttt{nicerarf} and \texttt{nicerrmf}, and the spectra were then binned using the ``optimal'' method (\citealt{kaastra2016}) in \texttt{ftgrouppha v1.6}.

The background that NICER experiences depends on several factors, including the location of the International Space Station in its orbit and the space weather conditions at the time of observation. The program \texttt{nibackgen3C50} (\citealt{remillard2022}) generates a background estimate by dividing the Good Time Intervals (GTI) from a given observation into smaller segments and then matching them to corresponding intervals from a library of reference spectra. The program uses different libraries depending on the “gain epoch” initially used to extract the spectrum, so for each observation we used ``\texttt{gainepoch=2020}'', which corresponded to the latest CALDB gain file at the time of our analysis.

Finally, the fully calibrated NICER pass band is generally taken to be 0.2--12.0 keV, but we have limited our energy range to 0.6-9.0 keV for the present analysis. The low-energy band was excluded because the detector flux calibration is less certain in this band, and we excluded the high-energy band because most of the individual spectra are dominated by background flux above 10~keV. An energy cutoff at 9~keV retains as much data as possible without background contamination adversely affecting the results of our analysis.

\subsection{Chandra}
Chandra observed NGC 4388 on two occasions, starting on 2008-04-16 02:03:04 (170.6~ks total exposure) and 2008-04-24 14:02:59	(98.3~ks total exposure), with the High Energy Transmission Gratings (HETG) in place.  We downloaded calibrated spectra and responses using the \texttt{tgcat} facility (http:$//$tgcat.mit.edu).  The exposures were obtained sequentially and are part of the same observation, so we co-added the first-order spectra from these observations using the CIAO tool \texttt{combine\_grating\_spectra}.  
After the processing applied in the \texttt{tgcat} sequencing, the net exposure time is 269~ks.

Our focus is on the detailed shape of the strong, neutral Fe~K$\alpha$ emission line, so we restrict our analysis to the summed first-order high energy grating (HEG) spectrum; it has the best spectral resolution and sensitivity in this band.  Prior to analysis, the data were binned using the HEASOFT tool \texttt{ftgrouppha}, again using the ``optimal'' binning algorithm.  Within XSPEC, the data were fit within the 5.0--7.5~keV band to ensure that our results are driven by aspects of the Fe~K$\alpha$ line. A Cash statistic (\citealt{cash1979}) was minimized to derive the best overall models and $1\sigma$ confidence limits.

\section{Analysis and Results \label{sec:analysis}}
\subsection{Spectral Modeling \label{sec:model}}
\begin{figure*}
    \centering
    \includegraphics[width=0.50\textwidth]{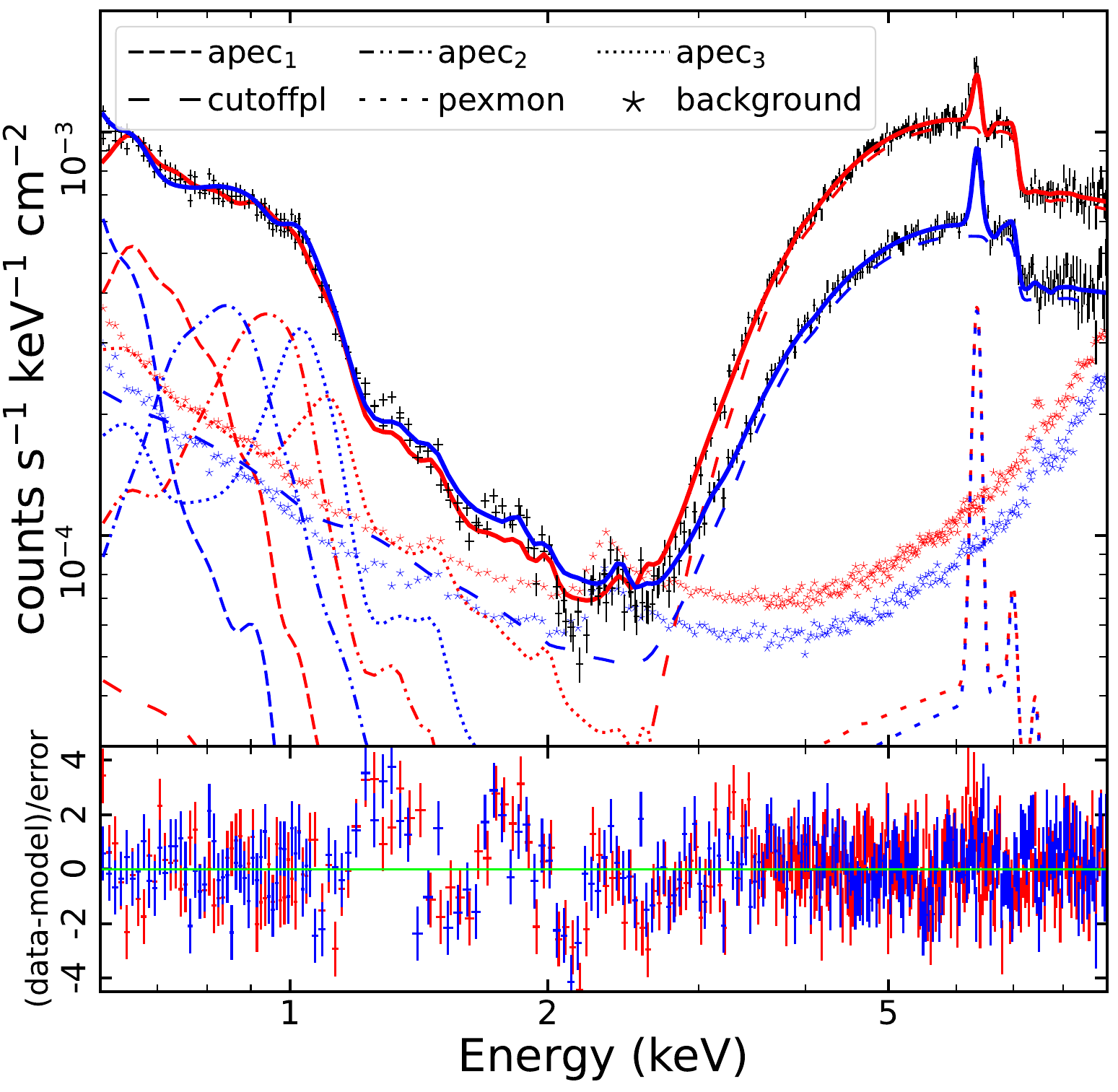}
    \hspace{-0.5em}
    \includegraphics[width=0.483\textwidth]{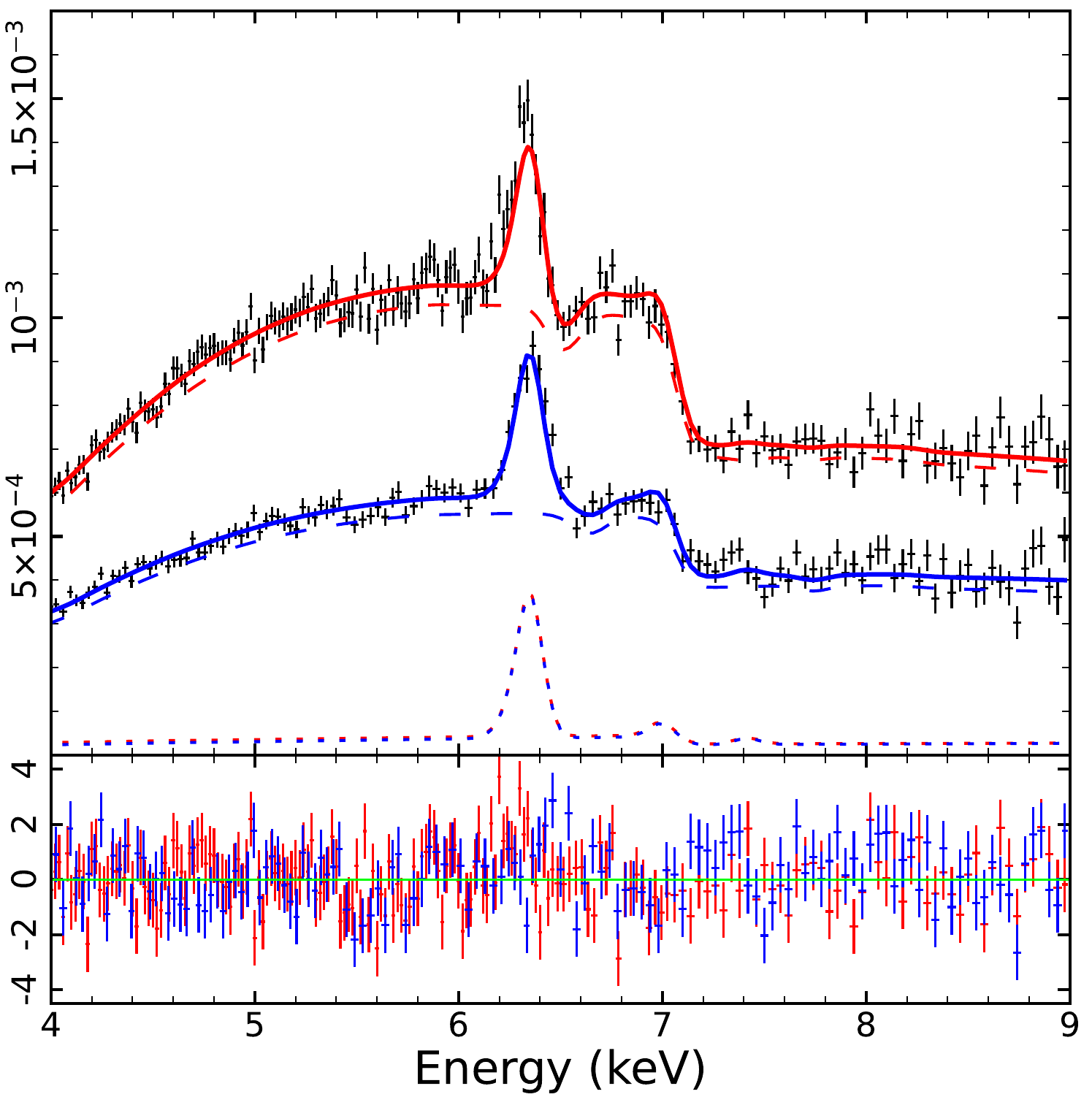}
    \caption{Time-averaged spectra of NGC 4388 from NICER. Both sets of data are individually fit to the model as described in Section \ref{sec:model}, with the best fit for the early epoch (2017-18) shown in red and for the late epoch (2022) shown in blue. Left: Full spectral fits from 0.6--9.0 keV. For each epoch, the total best fit model is shown with a solid line, and individual additive model components are shown with different line styles, as indicated in the legend. Background estimates generated by \texttt{nibackgen3C50} are plotted with stars. Right: The power law-dominated parts of both spectra shown from 4.0--9.0 keV, repeated here to highlight the Fe K$\alpha$ line in each spectrum as well as a small absorption feature captured by \texttt{ionabsorb}, which is more prominent in the early epoch. Both plots have been rebinned for visual clarity.}
    \label{fig:timeaverages}
\end{figure*}

\begin{figure*}
    \centering
    \includegraphics[width=1.0\textwidth]{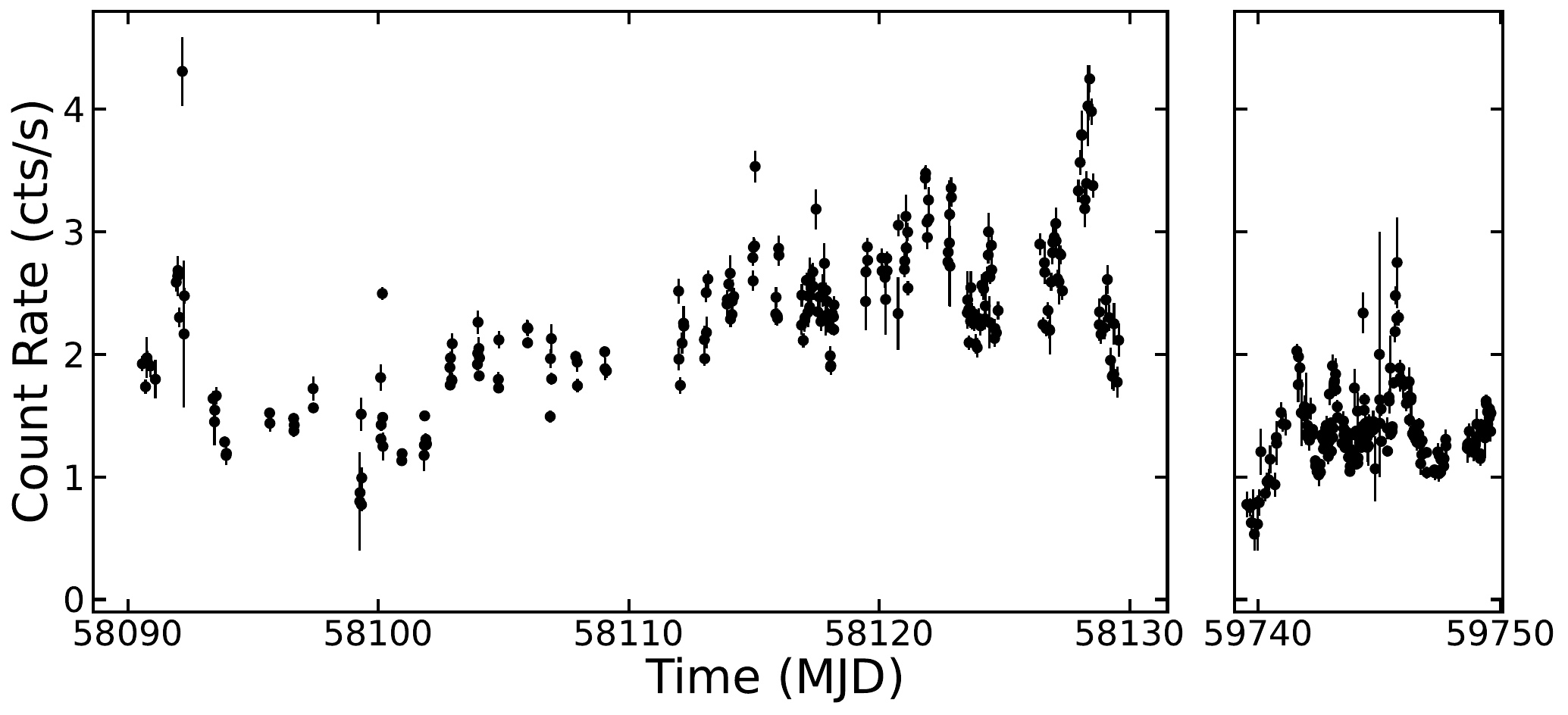}
    \caption{NICER light curve of NGC 4388 in the energy range 3-10 keV, indicative of the behavior of the power law continuum and reflection components of the X-ray spectrum. Each bin represents the average count rate over a period of 1000 s. The two epochs of observations used in this analysis are shown in adjacent panels, with over four years separating them. Note the greater breadth of the first epoch of observations, which made it better suited for our investigation of day-timescale variability, as detailed in Section \ref{sec:lag}. }
    \label{fig:lightcurve_3to10}
\end{figure*}

All fitting to NICER spectra was performed in XSPEC v12.12.0 (\citealt{arnaud1996}) and minimized a $\chi^2$ fit statistic using standard model weighting.

The model used for our primary investigation is as follows:

\texttt{TBabs*(apec\textsubscript{1} + apec\textsubscript{2} + apec\textsubscript{3} + TBpcf * ionabsorb * cutoffpl + pexmon)}

\texttt{TBabs} (\citealt{wilms2000}) models neutral absorption within the Milky Way, which in all fits is fixed at the value $2.57\times 10^{20}$ cm$^{-2}$ resulting from using the FTOOL \texttt{nH} (\citealt{kalberla2005}). \texttt{APEC} models emission from collisionally ionized, diffuse gas (\citealt{apec2001}), and three components of different temperatures were required to produce a good fit of the low-energy continuum at $\lesssim$ 3.0 keV. The main continuum from the central engine is modeled by a cut-off power law \texttt{cutoffpl}, which is modified both by an ionized absorber \texttt{ionabsorb} and a partial covering neutral absorber \texttt{TPpcf}. The ionized absorber is an XSTAR table model (see below); its function in these spectra is to model an absorption feature slightly above the Fe K$\alpha$ line. Initial fits poorly constrained the cutoff energy for \texttt{cutoffpl}, so it was frozen at 100 keV in all subsequent fits. Finally, a neutral reflection spectrum is modeled using \texttt{pexmon} (\citealt{pexmon}), which includes self-consistent Fe and Ni K$\alpha$ and K$\beta$ lines. The most prominent reflection feature in these spectra, necessitating the inclusion of this component, is the narrow Fe K$\alpha$ line at 6.4~keV. The parameters in \texttt{pexmon} for power law photon index, normalization, and cutoff energy are tied to those of \texttt{cutoffpl}. For \texttt{cutoffpl}, \texttt{pexmon}, and all \texttt{apec} components, the model parameter for redshift is frozen at the global redshift of the galaxy ($z=8.4\times 10^{-3}$, \citealt{lu1993}).

The XSTAR table model was constructed using the ``xstar2xspec'' functionality within the suite.  We assumed solar abundances for all elements based on \cite{grevesse1996}, and a turbulent velocity of $v_{turb} = 300~{\rm km}~{\rm s}^{-1}$.  The input SED consisted of a hot, $T = 25,000$~K blackbody and $\Gamma = 1.7$ power-law continuum (artificially bent to zero at low energy) in a 10:1 flux ratio typical of bright, unobscured AGN. The ionizing luminosity was assumed to be $L = 1\times 10^{44}~{\rm erg}~{\rm s}^{-1}$ over the 0.0136--13.6~keV band. The grid points were selected to ensure at least 10 points per decade in the critical parameters: 80 points between $-2 < {\rm log}\xi < 6$ and 40 points between $1\times 10^{20}~{\rm cm}^{-2} < {\rm N}_{\rm H} < 6\times 10^{23}~{\rm cm}^{-2}$.  

After fits were made with this model, we calculated the power law flux by adding a \texttt{cflux} component to \texttt{cutoffpl}, such that the power law component of the model reads as \texttt{TBpcf * ionabsorb * cflux * cutoffpl}. We also calculated the reflected flux in a similar manner, using \texttt{cflux * pexmon} in separate fits. In both cases, the energy bounds of \texttt{cflux} were set to the full energy range of the present analysis, 0.6--9.0 keV, and all model parameters were then frozen except for the ``lg10Flux'' parameter of \texttt{cflux}. To facilitate better comparison with other measurements in the literature, we also repeated this process with energy bounds at 2--10 keV to calculate fluxes in this more standard energy range.

All uncertainties presented in this analysis reflect 1$\sigma$ confidence intervals calculated with the \texttt{error} command in XSPEC, unless otherwise specified. These values were corroborated with the results from running \texttt{steppar} as necessary, if non-monotonicity in the fit statistic was detected.

\subsection{Long-Term Variability \label{sec:longterm}}
Figure \ref{fig:timeaverages} and Table \ref{table:timeaverage_results} summarize the results of the best fits to the time-averaged spectra from each epoch of observations. Comparing the model parameters of both time-averaged spectra allows us to determine how the overall behavior of NGC 4388 changes on a timescale of approximately 1650 days (4.5 years). Figure \ref{fig:timeaverages} clearly reveals a significant difference in the flux detected above approximately 3~keV, compared to the relatively constant flux measured below 3~keV. While our modeling of the emission below 3~keV with \texttt{apec} does result in statistically significant variations in temperature and normalization, there are also a number of factors complicating our confidence in this result. Most noticeable among them are the large residuals between $\sim$1--2 keV, which lie in the range of known NICER detector features. As seen in the left panel of Figure \ref{fig:timeaverages}, the profiles of corresponding \texttt{apec} components between the two models are generally similar if shifted in energy slightly, and there is also a significant change between the two power law components of the larger model; the \texttt{cutoffpl} component contributes more flux in this energy band in the later than the earlier epoch. Along with the fact that the spectra become locally background-dominated around 2 keV, it is possible that detector features, small changes in the background estimation, or differences in the \texttt{cutoffpl} model component are contributing to different best-fit values for the various \texttt{apec} components.

\begin{deluxetable}{cc|cc}
    \tablecaption{Results of Time-Averaged Epoch Spectral Fitting}
    \label{table:timeaverage_results}
    \tablewidth{\columnwidth} 
    \tabletypesize{\scriptsize}
    \tablehead{
        \colhead{Component} & \colhead{Parameter} & \colhead{Early Epoch} & \colhead{Late Epoch}
    }
    \startdata
        \texttt{apec\textsubscript{1}} & kT (keV) & $0.282(5)$ & $0.188_{-0.006}^{+0.007}$ \\
        & norm ($10^{-4}$) & $1.93_{-0.09}^{+0.07}$ & $2.4(2)$ \\
        \texttt{apec\textsubscript{2}} & kT & $1.01_{-0.02}^{+0.01}$ & $0.75_{-0.02}^{+0.01}$ \\
        & norm & $1.61(5)$ & $1.11(5)$ \\
        \texttt{apec\textsubscript{3}} & kT & $2.9_{-0.8}^{+0.3}$ & $1.41(3)$ \\
        & norm & $5.5_{-2.1}^{+0.9}$ & $2.9(1)$ \\
        \texttt{TBpcf} & N$_{\text{H}} \; (10^{22} \; $cm$^{-2})$ & $40.3_{-0.6}^{+0.7}$ & $39.6_{-0.3}^{+0.5}$ \\
        & pcf & $0.999_{-0.002}^{+0.001}$ & $0.9868_{-0.0008}^{+0.0007}$ \\
        \texttt{ionabsorb} & N$_{\text{H}}$ & $2.4_{-1.5}^{+0.7}$ & $0.8(2)$ \\
        & $\log\xi$ & $3.2_{-0.1}^{+0.7}$ & $3.67_{-0.09}^{+0.22}$ \\
        \texttt{cutoffpl} & $\Gamma$ & $1.52(3)$ & $1.32_{-0.04}^{+0.03}$ \\
        & norm ($10^{-2}$) & $3.0(2)$ & $1.07(7)$ \\
        \texttt{pexmon} & rel\_refl & $-0.135_{-0.009}^{+0.008}$ & $-0.24(1)$ \\
        \hline
        \multicolumn{2}{c|}{$f_{\text{tot}}\;(10^{-11}\;$erg cm$^{-2}$ s$^{-1}$)} & 4.78(2) & 2.77(1) \\
        \multicolumn{2}{c|}{$f_{\text{PL,0.6--9.0}}$} & 19.9(5) & 9.0(2) \\
        \multicolumn{2}{c|}{$f_{\text{refl,0.6--9.0}}$} & $0.27_{-0.02}^{+0.01}$ & $0.25(1)$ \\
        \multicolumn{2}{c|}{$f_{\text{PL,2--10}}$} & $15.2(2)$ & $7.6(1)$ \\
        \multicolumn{2}{c|}{$f_{\text{refl,2--10}}$} & $0.31(2)$ & $0.29(1)$ \\
        \hline
        \multicolumn{2}{c|}{$L_{X,\text{2--10}}\;(10^{42}$ erg s$^{-1})$} & $5.91_{-0.07}^{+0.08}$ & $2.94(4)$ \\
        \multicolumn{2}{c|}{$L_{X,\text{2--10}}/L_{\text{Edd.}}$} & $0.00559(7)$ & $0.00278(4)$ \\
        \hline
        \multicolumn{2}{c|}{$\chi^2/\nu$} & $1023.53/825=1.24$ & $952.92/825=1.15$
    \enddata
    \tablecomments{Key model parameters from fits to time-averaged NICER spectra for NGC 4388. All reported errors are 1$\sigma$ uncertainties in the model parameters, and numbers in parentheses represent symmetric errors in the last digit quoted. \\ The low-end continuum is dominated by three \texttt{apec} plasma components, each with a characteristic temperature $kT$ and normalization. The main power law continuum \texttt{cutoffpl} has a photon index $\Gamma$ and normalization, and it is subject to two different absorbers. The neutral \texttt{TBpcf} has a column density $N_H$ and partial covering fraction $pcf$, and the ionized \texttt{ionabsorb} has an ionization parameter $\xi$ and separate column density. The reflection component \texttt{pexmon} shares its photon index and normalization with \texttt{cutoffpl}, so its strength is controlled by the reflection fraction $rel\_refl$. The reflection fraction is presented as a negative number, as it appears when fitting, because \texttt{pexmon} uses a minus sign to indicate that only the reflection has been modeled and not the power law and reflection together. We have chosen to model the continuum and reflection as separate components, hence the minus sign. \\ Also presented here are the total received flux $f_{\text{tot}}$, unobscured power law flux $f_{\text{PL,0.6--9.0}}$, and reflection component flux $f_{\text{refl,0.6--9.0}}$, along with adjusted values in the more standard range 2.0-10.0 keV, denoted $f_{\text{PL,2--10}}$ and $f_{\text{refl,2--10}}$. The 2--10 keV X-ray luminosity $L_{X,\text{2--10}}$ is calculated using $f_{\text{PL,2--10}}$ and a distance of 18.0 Mpc (\citealt{sorce2014}), and the Eddington fraction $L_{X,\text{2--10}}/L_{\text{Edd.}}$ is presented for reference (see Section \ref{sec:shortterm}).}
\end{deluxetable}

The column density of the neutral absorber \texttt{TBpcf} is consistent between both epochs, with both fits yielding a similar value of around $4\times10^{23}$ cm$^{-2}$. This value is higher than previous attempts at modeling these data (\citealt{miller2019}) but is in broad agreement with historical measurements of the column density of this source (\citealt{elvis2004}). The associated covering factors for each of these column densities are also very similar, so together both pairs of parameters are consistent with a neutral absorber whose average behavior does not change drastically over a period of several years, either in column density or its covering fraction toward our line of sight.

Despite this apparent constancy in the absorbing column, we still observe changes in continuum emission on these long timescales. This variability in power law flux---a change of a factor of two between epochs---therefore likely indicates true changes in the behavior of the central engine, rather than just changes in obscuration in our line of sight. As the flux of this part of the continuum decreases from the first to the second epoch, so too does its spectral slope; the photon index drops significantly from $\Gamma=1.52\pm0.02$ to $\Gamma=1.32_{-0.04}^{+0.03}$.
\begin{figure}[t]
    \centering
    \includegraphics[width=\columnwidth]{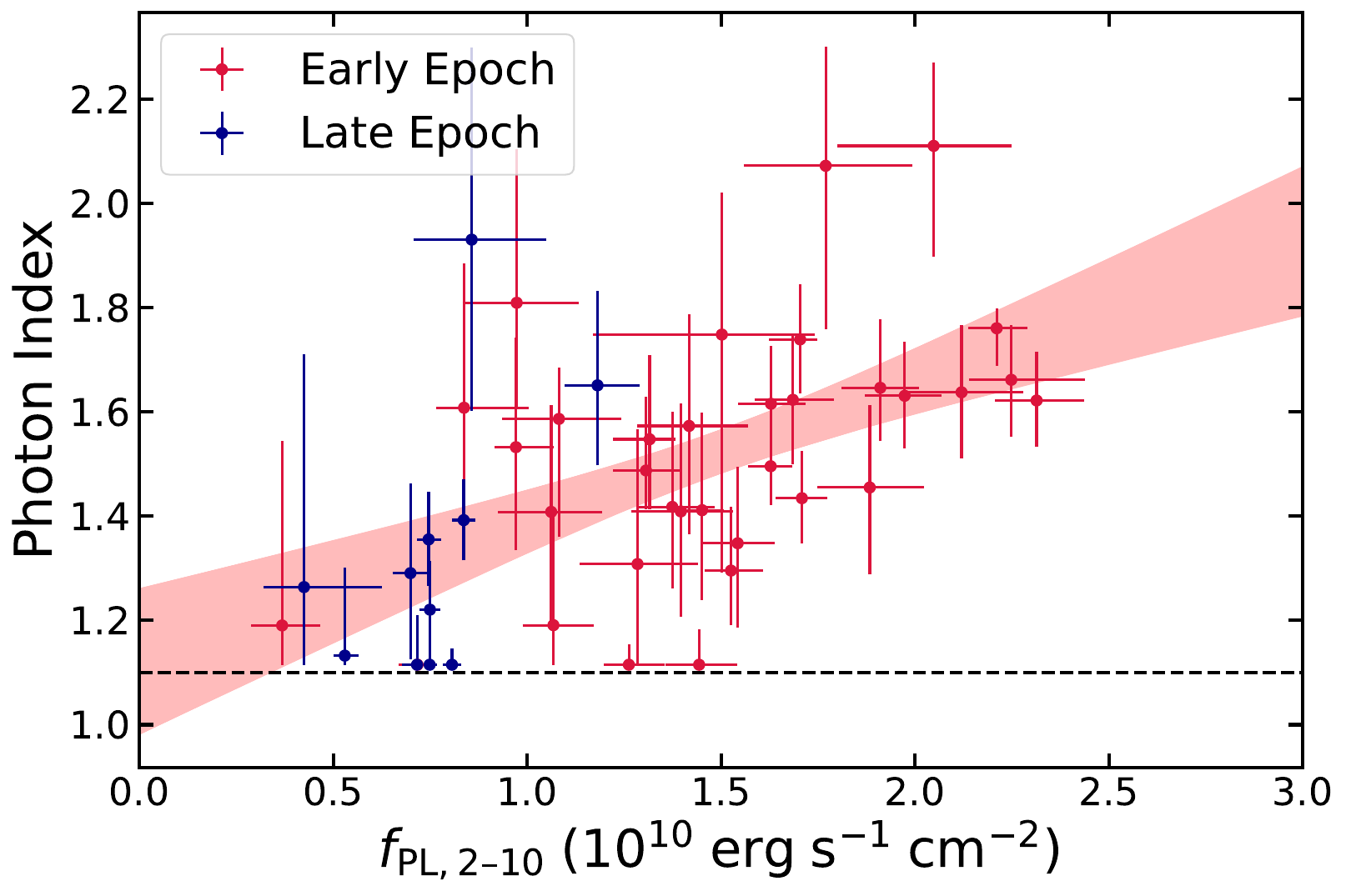}
    \caption{Photon index vs. unabsorbed power law flux, $f_{\text{PL,2--10}}$, for individual observations in both epochs. The black dashed line is $\Gamma=1.115$, indicating the hard lower bound imposed when fitting due to the limitations of \texttt{pexmon}. The pink shaded region shows sample fits from a Monte Carlo procedure, fitting both epochs jointly. The samples were drawn from the 68\% contour around the mean values for the best-fit slope and intercept, so this band represents a $1\sigma$ region of uncertainty.}
\label{fig:gamma_plot}
\end{figure}

\subsection{Short-Term Variability} \label{sec:shortterm}
Motivated by the spectral variability seen between these two observational epochs, we next focused on variability among the 45 individual observations that make up both epochs. The most drastic and most evident change between the two time-averaged spectra in Section \ref{sec:longterm} is in unobscured power law flux, which is an indicator of the behavior of the central engine. The magnitude of this observed variability therefore enables us to investigate how other parameters change with or respond to the central engine.

For sufficiently luminous AGN, we typically observe a ``softer-when-brighter'' phenomenon, in which an increasing proportion of the power law continuum comes from soft X-rays as total luminosity increases (\citealt{yang2015}). Physically, this can be explained by a more luminous central engine---emitting more photons per unit time---cooling the corona more effectively. Compton up-scattering by electrons in the corona is theorized to be most responsible for X-ray continuum emission, so a cooler corona would result in less emission in harder X-rays and a softer continuum overall (\citealt{fanali2013}).

This phenomenon can be observed by investigating the relationship between power law flux and spectral slope, $\Gamma$, which is shown in Figure \ref{fig:gamma_plot}. From fitting all individual observations to a line using a Monte Carlo fitting procedure, we find a best fit of $\Gamma=(0.27\pm0.06)\times f_{\text{PL,2--10}}+(1.11\pm0.09)$, using the same units as in the figure. Since a higher value of $\Gamma$ corresponds to a ``steeper'' spectral slope and lower flux at higher energies, this positive correlation does indeed indicate ``softer-when-brighter'' behavior in NGC 4388. This is expected for AGN with $L_X/L_{\text{Edd.}}>10^{-3}$ (\citealt{yang2015, connolly2016}), and NGC 4388 agrees with this trend in both epochs (see Table \ref{table:timeaverage_results}).

We note that some individual observations reached a hard lower limit of $\Gamma=1.1$ when fitting, which is the lowest value allowed by \texttt{pexmon}. This limit is the point below which calculation of the Fe and Ni K$\alpha$ and K$\beta$ lines breaks down, indicating more unphysical behavior. The majority of these offending observations suffer from low exposure times and/or excess background contamination, so these values of $\Gamma$ and their uncertainties are therefore less reliable and could affect the numerical results of the best fits. However, fitting with these observations removed still yields the same qualitative result of a positive correlation between photon index and power law flux; a simple Spearman's rank correlation test returns a correlation coefficient of $r_s=0.635$ ($p=3\times10^{-6}$) for all observations and $r_s=0.596$ ($p=6\times10^{-5}$) with the $\Gamma=1.1$ observations removed. We note that joint fits to INTEGRAL and Swift observations of NGC 4388 measured a power-law index of $\Gamma = 1.2$ in some flux and hardness windows \citep{fedorova2011}; it is possible that we have sampled similar phases.

\subsection{Lag Analysis \label{sec:lag}}
\begin{figure}[t]
    \centering
    \includegraphics[width=\columnwidth]{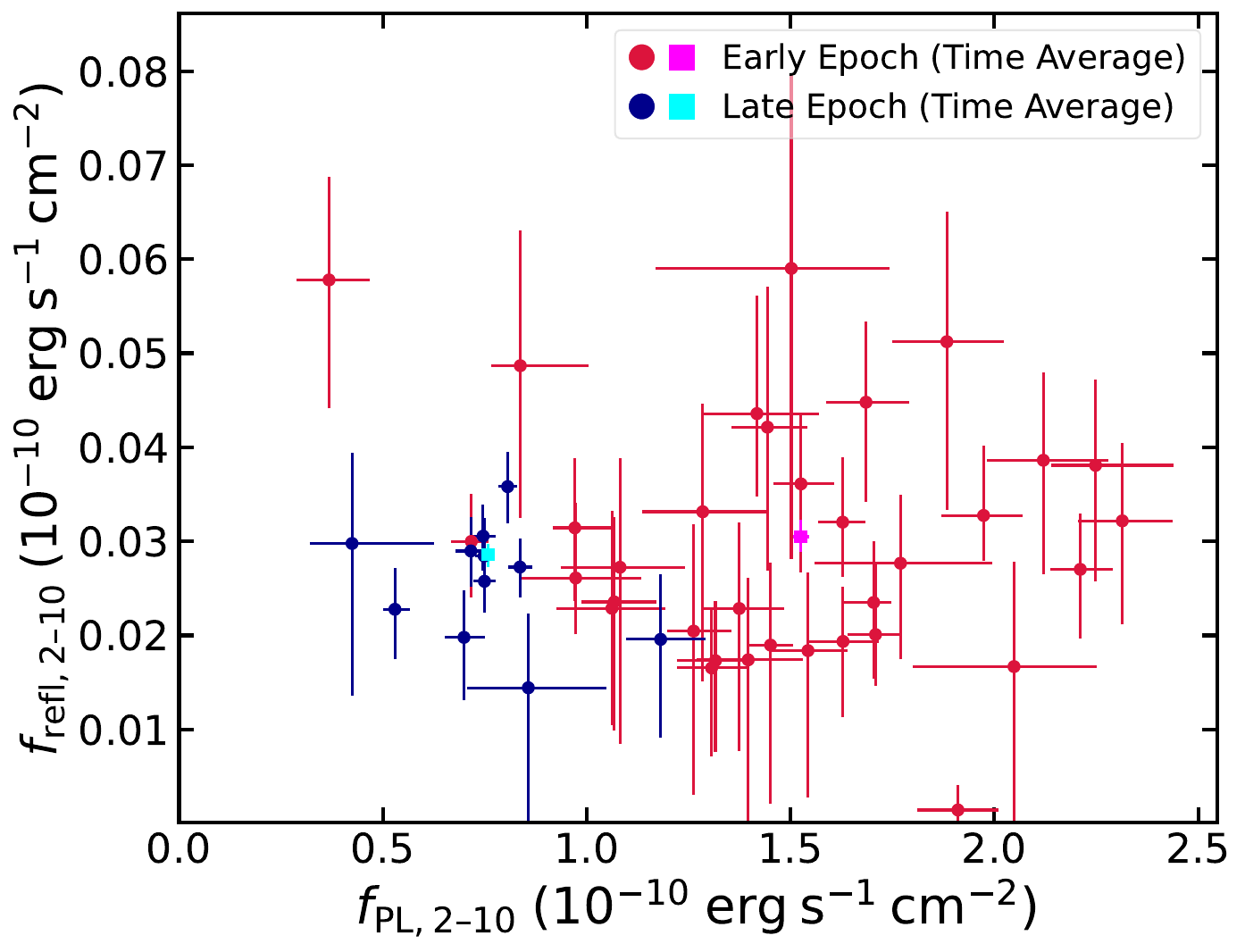}
    \caption{The reflected flux (modeled with \texttt{pexmon}) vs. the direct power-law continuum flux for NGC 4388. Individual observations of the first (second) epoch are shown in red (blue) and their time average spectrum is shown in magenta (cyan). Although the late epoch tends toward lower power law flux values (see Figure \ref{fig:timeaverages}), there is no strong correlation regardless of epoch.}
    \label{fig:fluxplot}
\end{figure}

After investigating the relationship between the flux of the power law component and its spectral index, we then analyzed the power law's relationship to the reflected continuum. Figure \ref{fig:fluxplot} shows the reflected flux from \texttt{pexmon} versus the power law flux from \texttt{cutoffpl}; there is evidently a wide spread in the data. This relatively weak correlation between reflected flux and direct flux could be the result of failing to account for light travel times between the compact central corona close to the black hole, and the primary reflector.  Lags of this kind are commonly measured between the central engine and BLR in optical studies of AGN (\citealt{peterson1993,cackett2021}), and recently in X-ray studies as well (\citealt{zoghbi2019, noda2023}).  To test this possibility, we next undertook a lag analysis using a package commonly employed to characterize lags in optical studies of Seyferts and quasars.

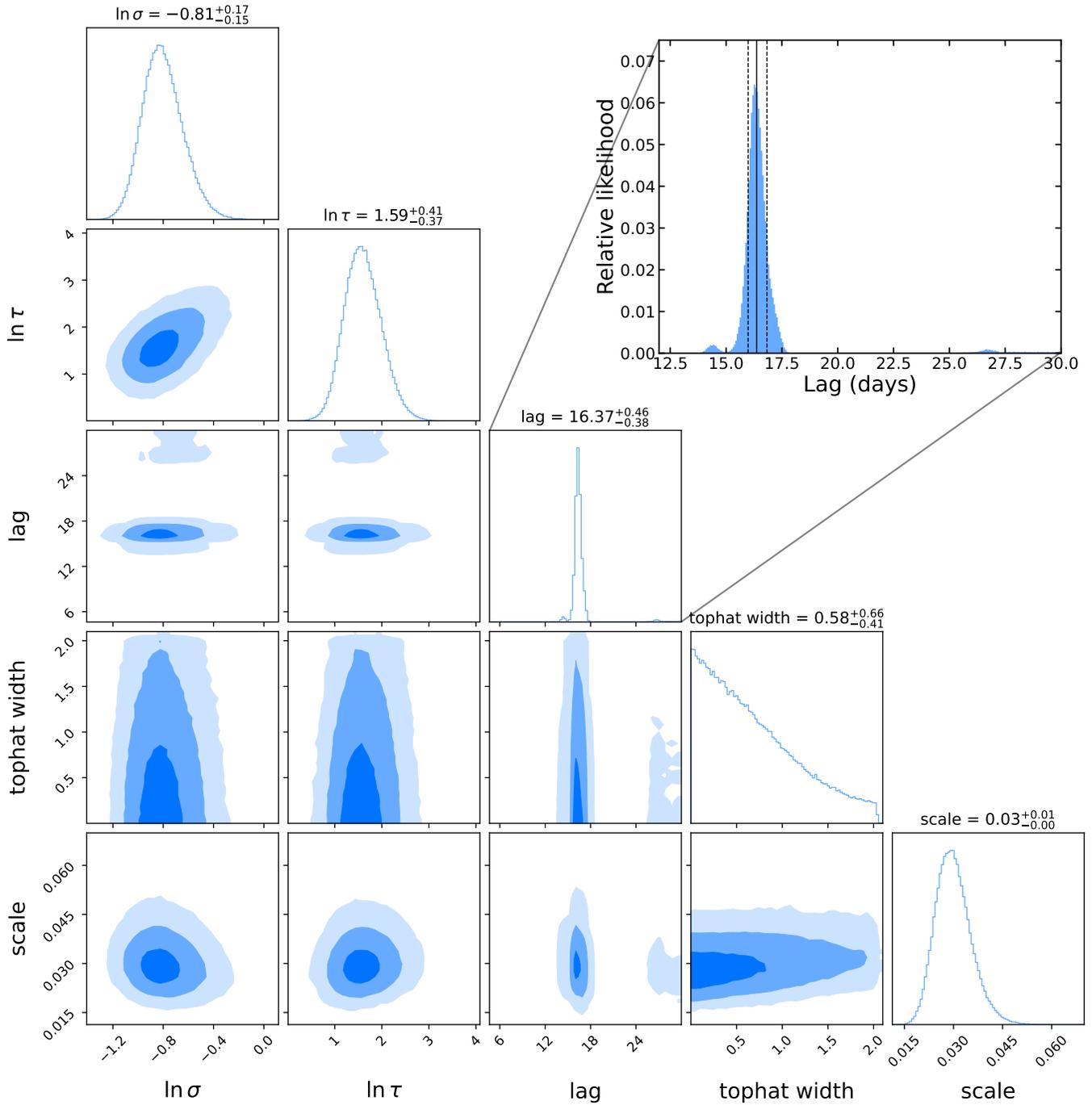
\begin{figure*}
    \centering
    \begin{tikzpicture}
        \node[anchor=south west, inner sep=0pt, outer sep=0pt] (corner) at (0,0){
            \includegraphics[width=\textwidth]{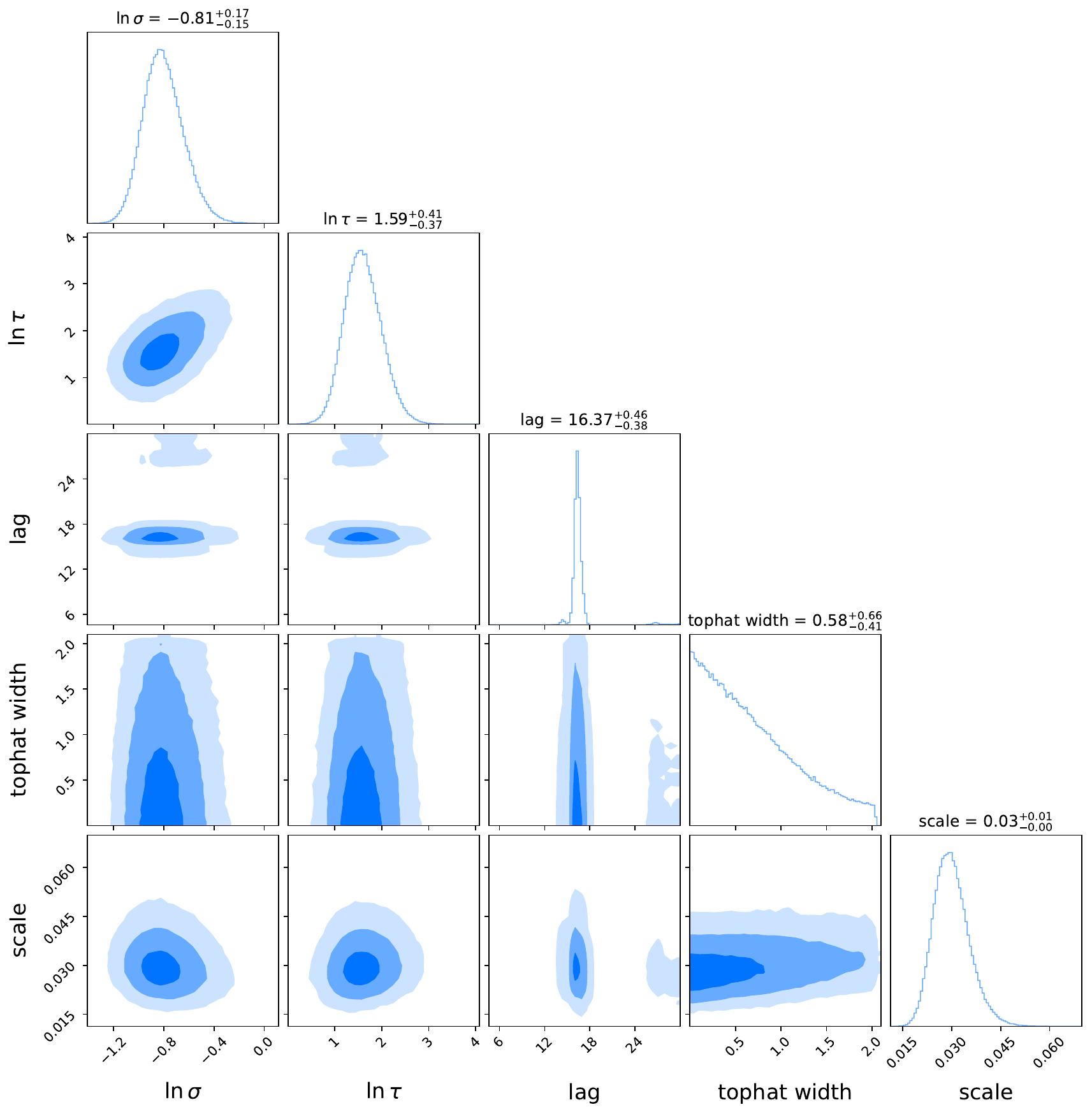}
        };
        \node[anchor=north east,xshift=-0.5em,yshift=-2em, inner sep=0pt, outer sep=0pt] (inset) at (corner.north east){
            \includegraphics[width=0.45\textwidth]{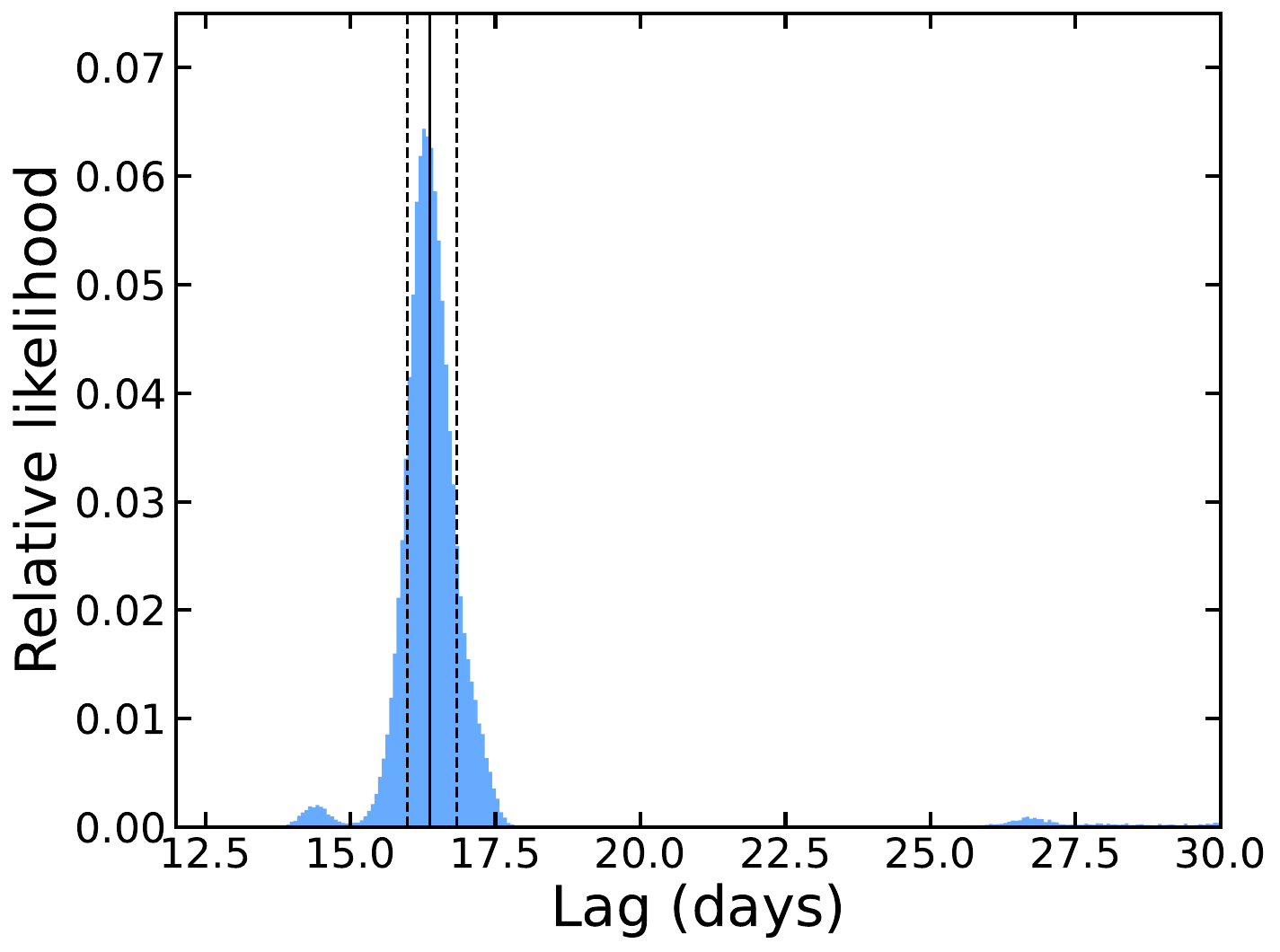}
        };
        \begin{scope}[x={(corner.south east)},y={(corner.north west)}]
            \coordinate (drawFromLeft) at (0.448,0.61);
            \coordinate (drawFromRight) at (0.623,0.439);
        \end{scope}
        \begin{scope}[shift=(inset.south west),x={(inset.south east)},y={(inset.north west)}]
            \coordinate (drawToLeft) at (0.137,0.985);
            \coordinate (drawToRight) at (0.95,0.13);
        \end{scope}
        \draw[thick,opacity=0.5] (drawFromLeft) -- (drawToLeft);
        \draw[thick,opacity=0.5] (drawFromRight) -- (drawToRight);
    \end{tikzpicture}
    \caption{Parameter distributions from a joint fit of the direct continuum and reflected flux light curves using JAVELIN. The light curves used encompass the first epoch of observation (see Figure \ref{fig:lightcurve}), and the increasingly lighter blue contours represent 1$\sigma$, 2$\sigma$, and 3$\sigma$ levels of confidence, respectively. The lag time, top-hat width, and $\tau$ are all in units of days, and the scale and $\sigma$ are unitless. The inset shows a more detailed view of the distribution of lag times found by JAVELIN; the vast majority of samples are located around a single peak. The solid black line shows the mean lag value, $t=16.37$ days, and the dotted lines bound a 1$\sigma$ credible interval.}
    \label{fig:cornerplot}
\end{figure*}

JAVELIN is a Python software package that computes lags between continuum and emission-line spectra (\citealt{zu2011}).  It is fully general and can be applied to X-ray data (see, e.g., \citealt{zoghbi2019}).   JAVELIN can estimate lags even within relatively sparse data by introducing two key assumptions about the nature of the light curves themselves.  First, JAVELIN assumes that a continuum light curve can be modeled as a damped random walk.  Second, JAVELIN further assumes that any emission-line light curve can be modeled as a shifted, scaled, and smoothed version of the continuum light curve.

\begin{figure}[b]
    \centering
    \includegraphics[width=\columnwidth]{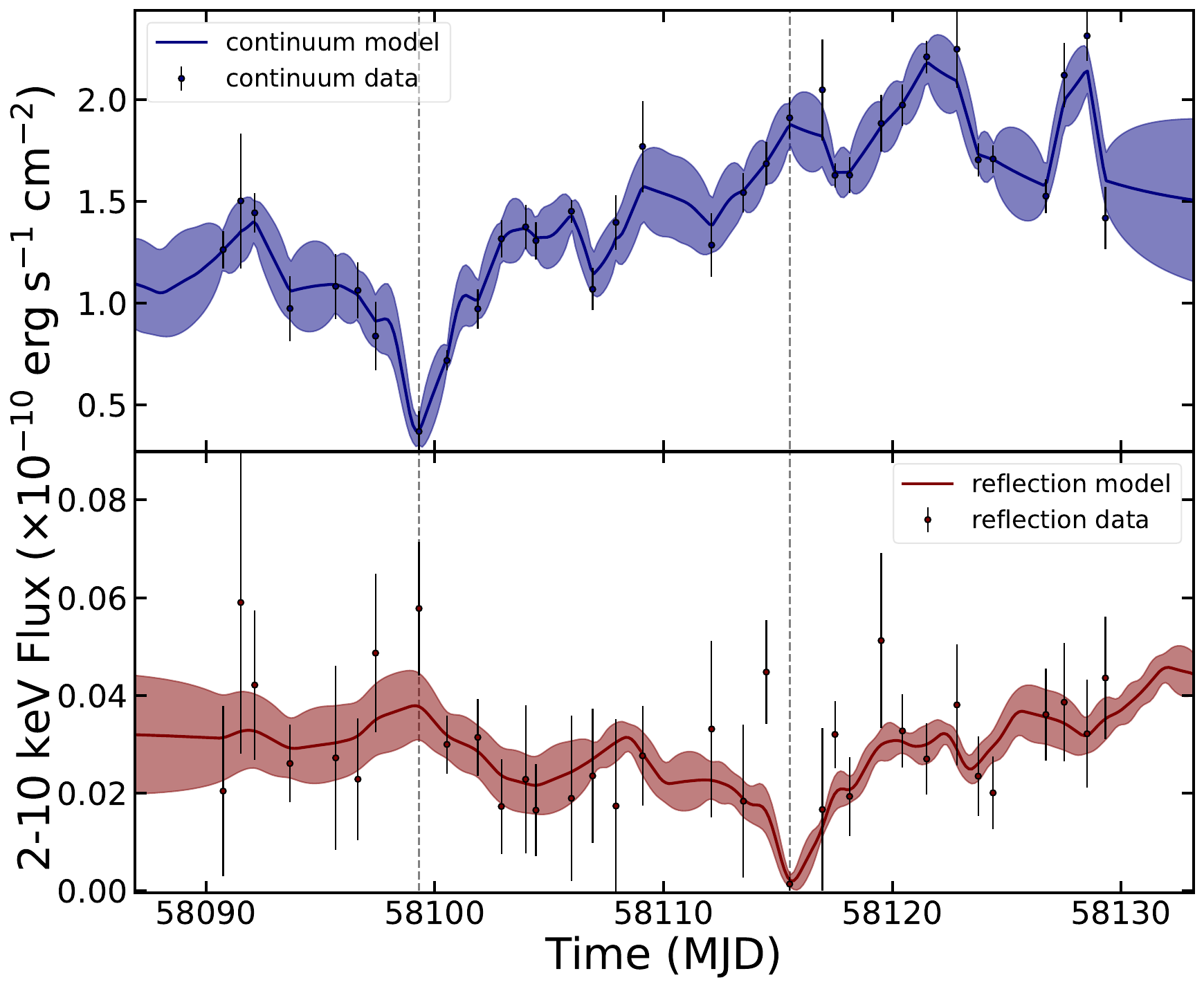}
    \caption{Light curves of NGC 4388 based on calculated power law flux ($f_{\text{PL,2--10}}$, top panel) and reflected flux ($f_{\text{refl,2--10}}$, bottom panel). All plotted points were derived using the \texttt{cflux} command in XSPEC on either the power law component of the model (\texttt{cutoffpl}) or the neutral reflection component (\texttt{pexmon}), which includes the neutral Fe K$\alpha$ line. The blue and red solid lines are the best-fit mean light curves based on fitting with JAVELIN, and the blue and red shaded regions represent regions of uncertainty around these best-fit models. The two dashed gray lines highlight the observations ObsID 1117010108 and 1117010121, whose spectra are presented in Figure \ref{fig:threeplot_thin}.}
    \label{fig:lightcurve}
\end{figure}

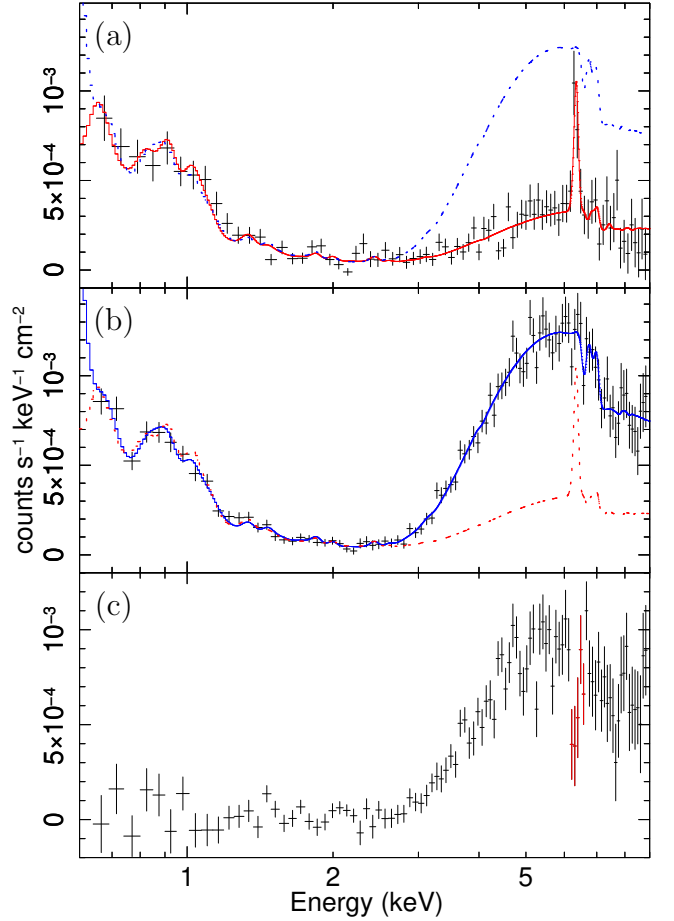
\begin{figure}[t]
    \begin{tikzpicture}
        \node[anchor=south west, inner sep=0pt] (image) at (0,0) {\includegraphics[width=\columnwidth]{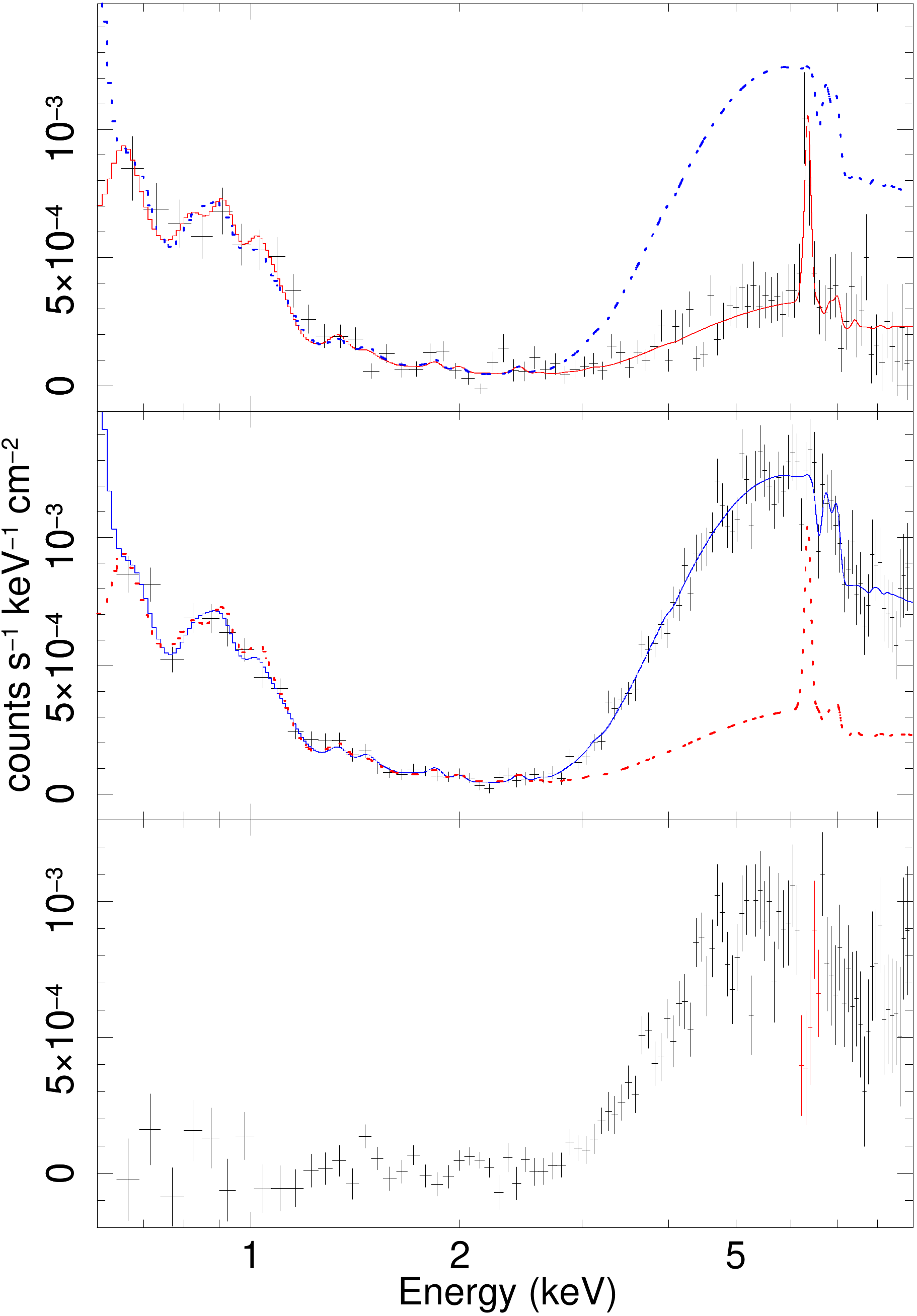}};
        \begin{scope}[
            x={(image.south east)},
            y={(image.north west)}]
            \node [black] at (0.16,0.96) {\large (a)};
        \end{scope}
        \begin{scope}[
            x={(image.south east)},
            y={(image.north west)}]
            \node [black] at (0.16,0.65) {\large (b)};
        \end{scope}
        \begin{scope}[
            x={(image.south east)},
            y={(image.north west)}]
            \node [black] at (0.16,0.34) {\large (c)};
        \end{scope}
    \end{tikzpicture}
    \caption{An example of short-term variability between two individual NICER observations of NGC 4388. (a) Spectrum of ObsID 1117010108 (MJD 58099), with the best-fit model shown in red. (b) Spectrum of ObsID 1117010121 (MJD 58115), with the best-fit model shown in blue. In both panels, the best-fit model of the other observation is shown with a dotted line for reference. (c) Difference spectrum of (b) and (a), with the range 6.1-6.7 keV highlighted in red. Note that the spectra are similar at $\lesssim$ 3 keV and differ substantially for $>3$ keV. The difference in time between the medians of these two observations is 16.20 days, comparable to the lag time found by JAVELIN.}
    \label{fig:threeplot_thin}
\end{figure}

\begin{figure*}
    \centering
    \includegraphics[width=\textwidth]{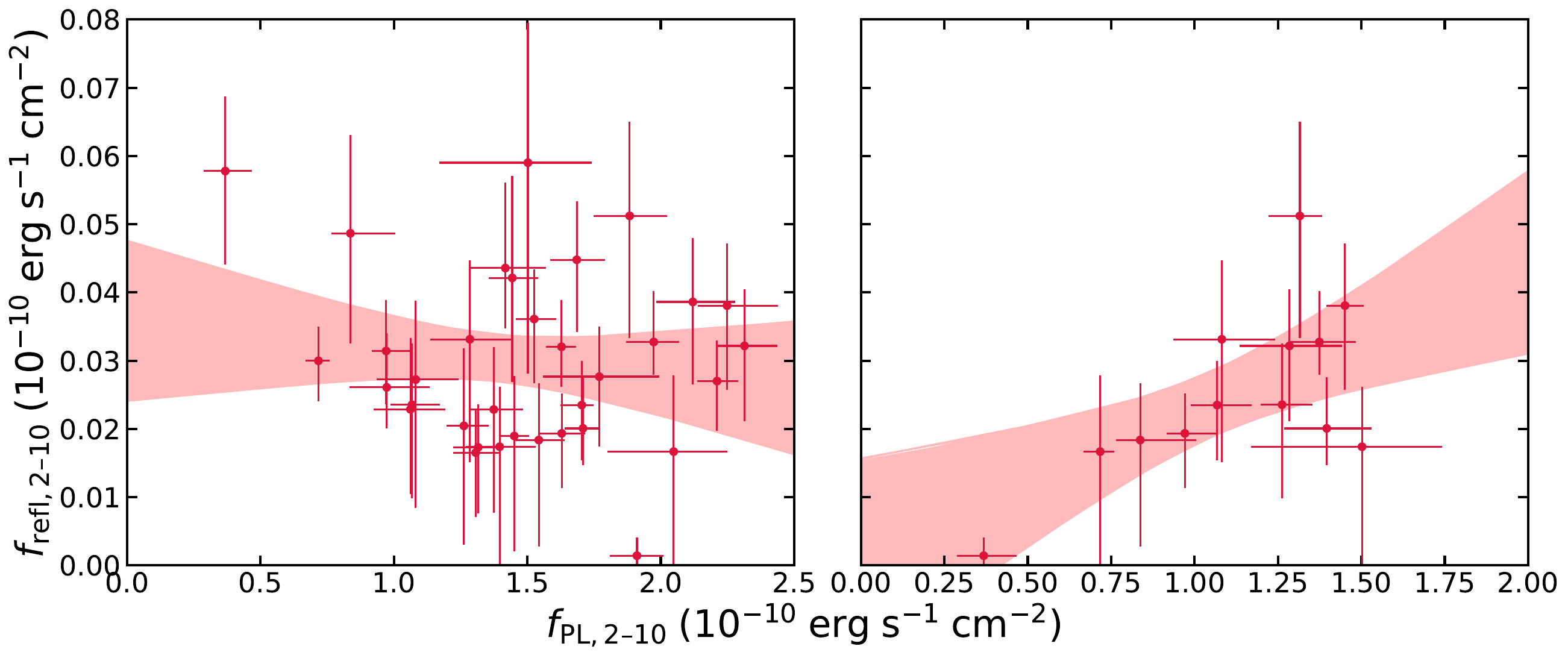}
    \caption{Fits to plots of reflected flux ($f_{\text{refl,2--10}}$) vs. power law flux ($f_{\text{PL,2--10}}$) for the first epoch of NICER observations. The pink shaded regions encompass sample fits drawn from the 68\% contour from Monte Carlo fitting procedures, representing an approximate $1\sigma$ region of uncertainty in the best fit in each panel. Left: Original data, where each data point corresponds to a distinct observation. The best fit to these data has a slope consistent with zero, indicating no correlation between the two quantities.  Right: Lag-adjusted data, where each data point corresponds to the power law flux of one observation and the reflected flux of a different observation taken within the range of $t=16.37_{-0.38}^{+0.46}$ days later. Because of the cadence of observing in this epoch, there are only 13 pairs of observations that meet these criteria. The best fit to these data has a slope of $m=(2.2\pm0.9)\times10^{-2}$, which indicates a positive correlation between power law flux and reflected flux after the characteristic lag time.}
    \label{fig:fluxplot_lag}
\end{figure*}

JAVELIN models this damped random walk with two parameters, one for a characteristic timescale ($\tau$) and one for a characteristic flux amplitude variability ($\sigma$).  This model is then convolved with a ``top-hat'' function to fit the line emission light curve, using three additional parameters: (1) a horizontal shift, describing a lag time between the continuum and line emission; (2) an amplitude, or a scaling factor between the magnitude of continuum and line flux; and (3) a top-hat width, effectively blurring the continuum signal within a given range.  The value of the lag time parameter is of primary interest to the present investigation, since a measure of the lag $t$ between a continuum signal and emission line yields a characteristic distance scale $d=ct$ for the line emitting region.  (In principle, the top-hat width parameter could also be used to constrain the extent of such a region, giving an inner and outer characteristic radius.)

To construct the light curves necessary for use as inputs to JAVELIN, we used the flux calculated from the \texttt{cutoffpl} and \texttt{pexmon} components, respectively, from the best-fit model of each individual observation in the first epoch of data collection. The \texttt{pexmon} component is dominated by the 6.4-keV iron line but also encompasses a more complex reflection spectrum. We use data from only the first epoch of observations because initial attempts using both epochs resulted in computed lag times on the order of the time between them. We therefore limited our scope to the available epoch with the longer period of observations. For each of these remaining 34 observations -- spanning a period of 39 days -- we calculated fluxes for the continuum and reflection using the method described in Section \ref{sec:model}. The associated time for each point on the light curve is the median of its observation time interval in MJD.

Figure \ref{fig:cornerplot} shows the resulting parameter distributions of an MCMC analysis in JAVELIN. The best-fit models resulting from these distributions are shown in Figure \ref{fig:lightcurve}, superimposed on the light curves used for the analysis. In JAVELIN, the argument \texttt{laglimit} was set to the range 0--30 days, to prevent negative lag times and to reduce the prevalence of calculated lags of length comparable to the total window of data collection. From these results, we find a lag time between continuum and reflection of $16.37_{-0.38}^{+0.46}$ days; if treated as light-travel time, and using a black hole mass of $M_{BH}=8.4\pm0.2\times10^6~M_{\odot}$ (\citealt{kuo2011}), this corresponds to a radius of $r=1.374_{-0.032}^{+0.039}\times10^{-2}\text{ pc }=3.4_{-0.1}^{+0.1}\times10^4\;GM/c^2$.

This derived lag time can be seen quite clearly by looking at the prominent dips in the modeled light curves of Figure \ref{fig:lightcurve}; there is a strong dip in the continuum close to MJD 58100 and a similar response by the reflection component some time later. The two individual observations closest to these two dips -- ObsIDs 1007010108 and 10007010121 -- are shown in Figure \ref{fig:threeplot_thin}. The time between the medians of both observations is 16.20 days, very similar to the lag found by JAVELIN. In the first observation, the power law continuum drops to significantly lower than average, but the Fe K line-dominated reflected continuum does not. The parameter $rel\_refl$ in \texttt{pexmon}, which is a scaling factor describing the relative strength of the reflected component compared to the direct continuum, has a magnitude of $0.99_{-0.28}^{+0.37}$ (note that the epoch average is $0.141_{-0.008}^{+0.008}$). In the second observation, the power law continuum returns close to the epoch average, but the Fe K line is very faint and almost appears to not be present; the magnitude of $rel\_refl$ can only be constrained with an upper bound of $0.028$. Even without relying on spectral modeling, the third panel of Figure \ref{fig:threeplot_thin} shows clear variability in the energy range corresponding to the power law continuum, as well as significant but distinct variability in the Fe K line. Between these two observations, then, we see strong variability in the direct continuum followed by strong variability in the reflected continuum, on a time scale consistent with that found by the JAVELIN analysis.

These two observations happen to have particularly strong changes in power law or reflected flux, and they seem to significantly affect JAVELIN's determination of a tentative lag timescale. Indeed, fitting the light curves from Figure \ref{fig:lightcurve} without these two observations present yields a much broader distribution of possible lags than those found in Figure \ref{fig:cornerplot}. Because of their strong effect on our variability analysis, we wanted to ascertain the significance of these two dip points within their respective light curves; our flux calculations are inherently model-dependent, so we wish to have some way of quantifying a level of confidence in the reliability of this lag determination and results.

In addition to fitting with JAVELIN, we also fit these light curves with a na\"{i}ve linear model to see by how much the flux values at the dip points vary from the average behavior of the rest of the light curve. In this sense, the power law flux at ObsID 1117010108 is lower than the model by $7.5\sigma$, and the reflected flux at ObsID 1117010121 is lower by $4.0\sigma$. We then included a Gaussian model component at the locations of the dips to find out by how much the normalization of the Gaussian excludes zero. Using this metric, the two dips are significant at the level of $12\sigma$ and $3.6\sigma$, respectively. In either case, these tests show that these two instances of variability are indeed significant; the fact that our estimation of a lag depends strongly on these points still means that it is strongly grounded in the data.

One additional check on the reliability of the results found by JAVELIN is understanding how adjusting for a lag of $t=16.37_{-0.38}^{+0.46}$ days improves the (lack of) correlation seen in Figure \ref{fig:fluxplot}. Figure \ref{fig:fluxplot_lag} presents fits to the reflected flux vs. power law flux data for the first epoch of observations, alongside points that have been adjusted by this tentative time delay. For example, the data point in the bottom-left of the right panel of the figure corresponds to the two observations in Figure \ref{fig:threeplot_thin}, showing the power law flux from ObsID 1117010108 and the reflected flux from ObsID 1117010121. Similarly, the other points in the panel correspond to any pairs of observations whose separation in time is within the $1\sigma$ errors of the JAVELIN lag. While fits to the original data show no correlation, as was evident in Figure \ref{fig:fluxplot}, fits to these lag-adjusted points have a best fit slope of $m=(2.2\pm0.9)\times10^-2$. This indicates a positive correlation between power law flux and reflected flux after a characteristic lag. This positive slope admittedly excludes zero at the level of $2.4\sigma$, but this still shows that this tentative lag is an improvement over the non-correlation of Figure \ref{fig:fluxplot} that first motivated our investigation.

\subsection{Fits to the Fe~K$\alpha$ Line at High Resolution \label{sec:chandra}}
For fits to the Fe K$\alpha$ line using Chandra data, we initially considered a simple model consisting of an absorbed, cut-off power-law and simple Gaussian line, \texttt{ztbabs*cutoffpl+zgauss}.  In this and all subsequent models, the redshift of the absorption, power-law continuum, and line was fixed to that of the host galaxy, $z = 0.008$.  The power-law cut-off energy was fixed at $E_{cut} = 100$~keV as per \cite{fabian2015}, and we fixed the Gaussian line energy at 6.400~keV.  Our results confirm those previously reported by \cite{shu2011}: we measure a line width of $FWHM = 2400\pm 600~{\rm km}~{\rm s}^{-1}$.  Assuming (a) that the broadening is due to Keplerian orbital motion and (b) that we observe the accretion flow at an inclination of $\theta = 60^{\circ}$, the line broadening would imply a radius of $r = 6.4^{+4.7}_{-2.3}\times 10^{4}~GM/c^{2}$.  The assumed inclination angle is broadly appropriate for Seyfert-2 AGN, but also the most probable angle at which a plane is likely to be viewed in 3D, and so a useful point of reference.

\begin{figure}[b]
    \centering
    \includegraphics[width=\columnwidth]{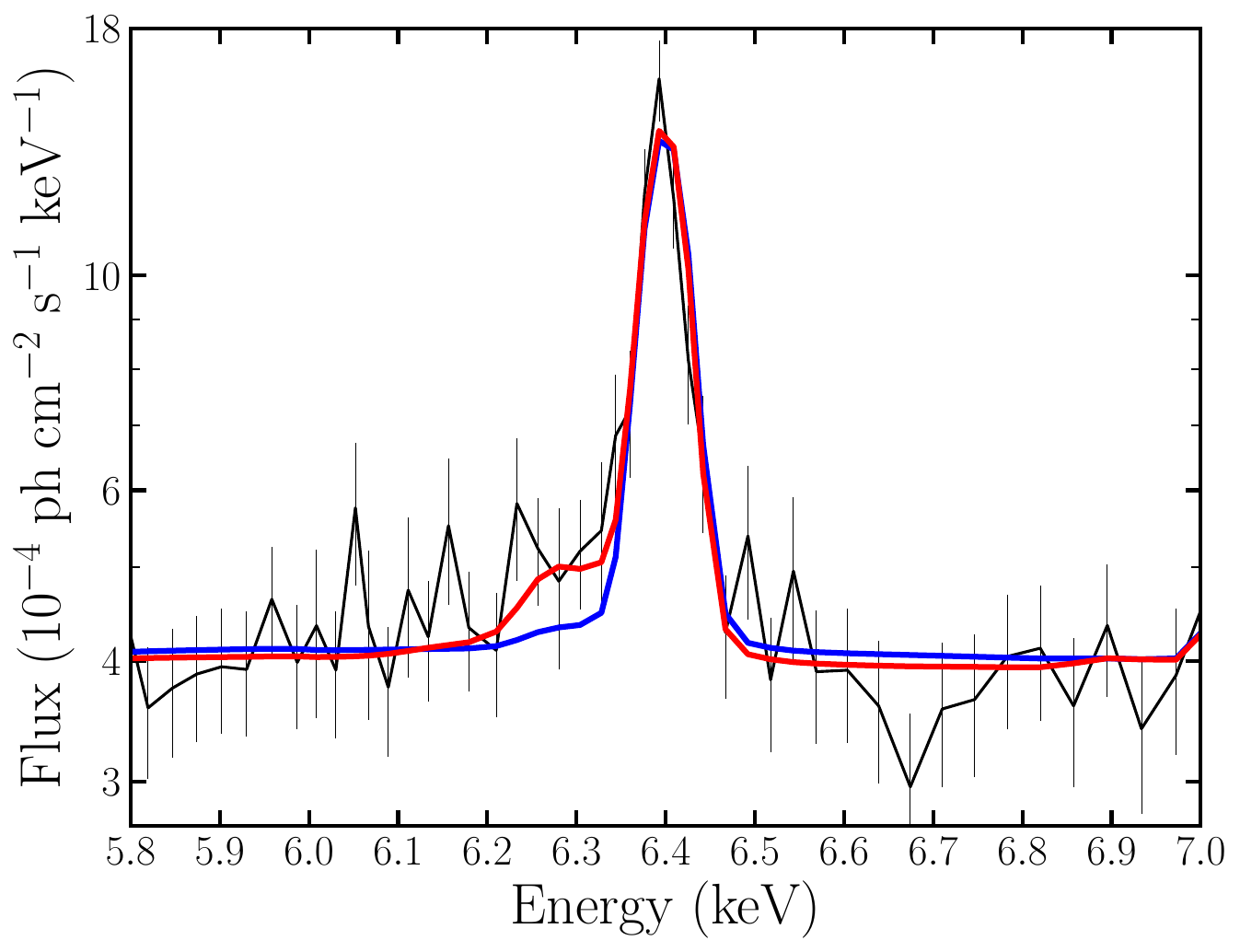}
    \caption{The summed Chandra HEG spectrum of NGC 4388 in the Fe~K band, shifted to the frame of the host galaxy.  The data represent 269~ks of net exposure and were binned using the ``optimal'' scheme of \cite{kaastra2016}, then binned additionally here for visual clarity.  The models are blurred line functions describing power-law irradiation of cold gas, \texttt{rdblur * mytorus}.  A fit with \texttt{mytorus} appropriate for Compton-thin gas ($N_{H} = 1\times 10^{23}~{\rm cm}^{-2}$) is shown in blue.  A superior model appropriate for Compton-thick gas ($N_{H} = 3\times 10^{24}~{\rm cm}^{-2}$) is shown in red; it provides a better fit to the Compton shoulder on the Fe~K$\alpha$ line that extends to 6.25~keV.}
    \label{fig:chandra}
\end{figure}

\begin{deluxetable*}{c|ccccccccc}
    \tablecaption{High-resolution Chandra Results}
    \label{table:chandra_results}
    \tablewidth{\textwidth} 
    \tabletypesize{\scriptsize}
    \tablehead{
        \colhead{Model} & \colhead{$N_{H}$} & \colhead{$\Gamma$} & \colhead{K} & \colhead{$R_{in}$} & \colhead{$N_{H,refl}$} & \colhead{$f_{refl}$} & \colhead{Norm.} & \colhead{C-stat} & \colhead{dof} \\
        \colhead{--} & \colhead{$10^{23}~{\rm cm}^{-2}$} & \colhead{--} & \colhead{$10^{-2}$} & \colhead{$10^{4}~GM/c^{2}$} & \colhead{$10^{24}~{\rm cm}^{-2}$} & \colhead{--} & \colhead{$10^{-2}$} & \colhead{--} & \colhead{--}
    }
    \startdata
        mytorus & 4.1(7) & 1.6(2) & $1.3^{+1.0}_{-0.5}$ & $2.9^{+1.2}_{-0.7}$ & 3   &    --   & $2.4^{+2.0}_{-1.1}$ & 147.2 & 159 \\
        mytorus & 4.2(7) & 1.6(2) & $1.4^{+1.0}_{-0.6}$ & $2.9^{+1.1}_{-0.7}$ & 1   &    --   & $1.9^{+1.5}_{-0.8}$ & 148.6 & 159 \\
        mytorus & 4.4(7) & 1.7(2) & $1.6^{+1.0}_{-0.6}$ & $2.6^{+1.0}_{-0.6}$ & 0.3 &    --   & $2.5^{+1.9}_{-1.0}$ & 156.2 & 159 \\
        mytorus & 4.6(7) & 1.7(2) & $1.7^{+1.2}_{-0.7}$ & $2.5^{+1.0}_{-0.7}$ & 0.1 &    --   & $2.4^{+0.9}_{-0.6}$ & 165.2 & 159 \\
        pexmon  & 4.5(7) & 1.7(2) & $1.4^{+1.0}_{-0.8}$ & $2.7^{+0.9}_{-0.5}$ & --  & -0.8(1) &         --          & 154.2 & 159
    \enddata
    \tablecomments{Spectral fits to the Fe K region of the combined first-order Chandra HEG spectra of NGC 4388.  The data were binned using the ``optimal'' binning scheme of \cite{kaastra2016} and fit over the 5.0--7.5~keV range.  The line was alternately described using the \texttt{mytorus} model or \texttt{pexmon}.  The former allows the column density of the line-emitting gas to vary while the latter assumes Compton-thick reflection.  Both models were blurred using \texttt{rdblur} assuming an isotropic point source emissivity of $J\propto r^{-3}$ and an inclination of $\theta = 60^{\circ}$. The $K$ parameter is the flux normalization of each model; in both cases, this is the flux normalization of the power-law, ${\rm ph}~ {\rm cm}^{-2}~ {\rm s}^{-1}~ {\rm keV}^{-1}$ at 1~keV.  The models all agree on an inner line emission radius that is formally consistent with the lag radius found by JAVELIN, and prefer a solution with Compton-thick emitting gas.  Please see the text for additional details.}
\end{deluxetable*}

The single Gaussian modeling of the Fe~K emission line is basic; the feature is actually composed of two lines.  Moreover, even two Gaussians would fail to account for the possibility of a Compton shoulder on the line, owing to $180^{\circ}$ scattering (this feature extends down to 6.25~keV; for a review, see \citealt{gallo2023}).  Moreover, even if the line is produced at distances of $r\simeq 10^{3-4}~GM/c^{2}$, special relativistic distortions to the line can still be important.  We therefore considered fits with \texttt{pexmon}, which self-consistently describes Compton-thick reflection from neutral gas, including Fe~K$\alpha$ and Fe~K$\beta$ lines as doublets, and with \texttt{mytorus} \citep{murphy2009}.  The latter model is a table calculated at very high (2~eV) resolution, wherein the column density of the line-emitting gas is a variable parameter. This parameter is sensitive to the relative strength of the line core and Compton shoulder.  We used the \texttt{mytorus} line table calculated for a power-law with a cut-off energy of $E_{cut} = 100$~keV.

Closely following fits to the line profile in Chandra observations of NGC 4151 (\citealt{miller2018}), we allowed for broadening by convolving \texttt{pexmon} and \texttt{mytorus} with \texttt{rdblur} (\citealt{fabian1989}).  This is an analytic blurring function that self-consistently accounts for relativistic Doppler shifts and gravitational redshifts around a black hole.  It explicitly assumes a zero-spin black hole, which is unlikely to describe any astrophysical black hole, but the potential far from a spinning black hole is very similar to the zero-spin case.  The four parameters of the \texttt{rdblur} model are the line emissivity (we assumed $R^{-3}$ as per an isotropic point source), the inner line-production radius (allowed to vary freely in our fits), the outer line-production radius (arbitarily fixed at $r_{out} = 10^{6}~GM/c^{2}$), and the inclination angle (assumed to be $\theta=60^{\circ}$ in all fits).  The XSPEC formulae for the two models are as follows: 
\texttt{TBabs*cutoffpl+rdblur*pexmon} and \texttt{TBabs*cutoffpl+rdblur*mytorus}.

The results of our fits with these models are detailed in Table \ref{table:chandra_results}.  They include one fit with \texttt{pexmon} and a series of four \texttt{mytorus} models that sample Compton-thin and Compton-thick gas.  The most important result from these fits is that all measurements suggest a line production radius of approximately $r = 3\times 10^{4}~GM/c^{2}$, with approximately 30\% uncertainties.  The best overall model is the \texttt{mytorus} fit that assumes Compton--thick gas with $N_{H} = 3\times 10^{24}~{\rm cm}^{-2}$; it measures a line production radius of $r = 2.9^{+1.2}_{-0.7}\times 10^{4}~GM/c^{2}$.   This model is better than that assuming Compton-thin gas, with $N_{H} = 1\times 10^{23}~{\rm cm}^{-2}$, by $\Delta{\rm C} = -18$ for the same number of degrees of freedom.  A comparison of these two particular models is shown in Figure \ref{fig:chandra}.  While our models effectively capture the details of the atomic features in the Fe~K band, it must be noted that the neutral absorption is treated in a simplified manner; the associated column densities should be regarded as fiducial and should not be contrasted with proper fits to the full pass band.

Finally, we included the ionized absorber model that was fit to the time-averaged NICER spectra, to test if this has any effect on the properties of the Fe~K$\alpha$ emission line modeling.  We find that the inclusion of an ionized absorber has no impact on the line properties derived using a Gaussian, \texttt{pexmon}, or \texttt{mytorus}.  The best-fit values and confidence intervals are entirely consistent.  The best-fit absorber was measured to have a column density of $N_{H} = 2.7^{+1.1}_{-1.1}\times 10^{22}~{\rm cm}^{-2}$, an ionization of ${\rm log}\xi = 3.6^{+0.1}_{-0.3}$, and a velocity shift of $v/c = -0.003^{-0.006}_{+0.005}$.  Relative to the host, this signals an outflow velocity of $v = -3300^{-1800}_{+1800}~{\rm km}~{\rm s}^{-1}$.
This nominally signals that the ionized absorption in NGC 4388 is indeed an outflow.  However, the absorber only improves the fit by $\Delta\chi^{2} = 12$ for $\nu = 3$ degrees of freedom, signaling that it is not highly significant.

\section{Discussion} \label{sec:discussion}
We have analyzed NICER monitoring observations of the nearby Seyfert-2 AGN, NGC 4388.  The two densest monitoring periods were selected and the summed spectrum from each was analyzed in detail.  We find no evidence for strong variations in the column density or covering factor of neutral obscuration along our line of sight to the central engine.  However, we measure significant changes in both the direct and reflected flux, and in highly ionized absorption that is evident in the Fe~K band.  In single monitoring observations, only a weak correlation is observed between the reflected and direct flux.  A search for lags between the direct and reflected flux reveals a potential lag of $t = 16.37^{+0.46}_{-0.38}$~days; a simple conversion of this lag to distance gives $r=1.374_{-0.032}^{+0.039}\times10^{-2}$~pc$\,=3.4\pm 0.1 \times 10^{4}~GM/c^{2}$, consistent with the optical BLR in unobscured AGN.  In this section, we discuss the strengths and weaknesses of our analysis and results, their implications, and plausible future observations that can build on these results.

The total spectral model that we have fit to the summed spectra from the first and third epochs (see Table \ref{table:timeaverage_results}) and the individual spectra within these monitoring periods is based on fits made to other highly absorbed Seyfert-2 AGN (see, e.g., \citealt{kammoun2020}).  It is likely that the soft X-ray spectrum is at least partly comprised of photoionized emission, whereas our modeling assumes collisional plasma emission.  However, for the purposes of this analysis, the simple \texttt{apec} components allow the flux to be characterized well, facilitating the detection of variations in the spectrum above 3 keV.  

Our spectral model also includes a component that measures highly ionized X-ray absorption in the Fe K band, following the detection of ionized Fe K absorption lines with, e.g., Suzaku and NICER (\citealt{shirai2008}, \citealt{miller2019}).  This absorption is only evident in the summed spectra from the first and third epochs; spectra from individual exposures lack the required sensitivity.  It is likely to be a low-velocity X-ray ``warm absorber.''   Utilizing the ionization parameter formalism, $\xi = L/nr^{2}$, assuming a bolometric correction of 15 (broadly consistent with \citealt{vasudevan2007}), and a unity filling factor so that $nr^{2} = Nr$, fits to the first and third epochs imply radii of $r \leq 0.92_{-0.27}^{+0.11}$ pc ($=2.3_{-0.7}^{+0.3}\times 10^6\;GM/c^2$) for the former and $r \leq 0.052_{-0.037}^{+0.029}$ pc ($=1.3_{-0.9}^{+0.7}\times 10^5\;GM/c^2$) for the latter.  These large radii are broadly consistent with the lack of a significant outflow velocity; however, the gas could originate at much smaller radii if the filling factor is small.  We note that the observed spectra have flatter power-law indices than was assumed by our photoionization table models, leading to systematic errors that are likely small compared to the uncertainties on the filling factor.  Future fits with the ``pion'' photoionization package can deliver more accurate results (\citealt{miller2015}, \citealt{mehdipour2016}).

Fits to the spectra from individual observations with the models developed for the summed spectra yielded useful results.  A significant positive correlation is observed between the power-law photon index, $\Gamma$, and the 2-10 keV power-law flux.  This correlation is typical of Seyfert-1 AGN, but has not previously been observed in NGC 4388 because its strong nuclear obscuration made it difficult to study the direct continuum in prior monitoring efforts.  Only a weak correlation is found between the reflected flux and direct power-law flux in the individual exposures; Figure \ref{fig:fluxplot} shows that a significant level of scatter is present in the relevant data.  This suggests that light travel times may not be properly accounted for, leading us to utilize JAVELIN to systematically search for lags between the direct continuum and reflection components.

The $t = 16.37^{+0.46}_{-0.38}$~day lag between the direct continuum and reflected flux measured with JAVELIN is partially driven by a sharp drop in the continuum flux around MJD 58099 (see Figure \ref{fig:lightcurve}).  The delayed dip is less pronounced in the reflected flux.  However, Figure \ref{fig:threeplot_thin} compares individual spectra obtained on MJD 58099 and MJD 58115, the two observations corresponding to these dips.  In the former, the line flux is unusually strong owing to the drop in the direct continuum; in the latter, the Fe K line is not evident at all.  This indicates that JAVELIN likely identified a delay time that is strongly grounded in the data.  However, this lag must be regarded as tentative, because the monitoring period is only twice the lag time.  Moreover, we have pragmatically utilized a simple, neutral reflector and applied it over a limited pass band, though some prior observations find evidence of ionized reflection in NGC 4388 (\citealt{risaliti2002}).  Future NICER observations that monitor over a period that is several times the duration of the tentative lag are required to confirm it.  Corresponding observations with NuSTAR will help to constrain the reflection.

In the meantime, current high resolution X-ray spectroscopy has provided one independent check.  Fits to archival high-resolution Chandra spectra of NGC 4388 measure line broadening consistent with a radius of $r = 6.4^{+4.7}_{-2.3}\times 10^{4}~GM/c^{2}$ for realistic assumptions.  This is nominally a factor of two larger than indicated by the tentative lag, though the $1\sigma$ error is quite close to the radius indicated by the tentative lag.  More physical models for the Fe~K region suggest line-production radii that are formally consistent with that indicated by the tentative lag; the best-fit model in Table \ref{table:chandra_results} measures $r = 2.9^{+1.2}_{-0.7}\times 10^{4}~GM/c^{2}$.  

The physical models also prefer that the line-emitting gas be Compton-thick.  If this is robust, it indicates that the line is produced through X-ray reflection as classically envisioned.  It is nominally easiest to associate the line-production region with the cold, optically thick accretion disk.  The fact that the fits pick out a specific radius suggests that the profile of the disk may not be locally flat, but rather have some vertical extent.  It is possible that a warp could provide the extra solid angle to anchor the narrow Fe~K line at a specific radius.   However, this scenario is likely also consistent with a wind that is infused with cold, optically thick clumps.  One model of the optical BLR posits that it is a wind that is initially launched by radiation pressure on iron--rich dust, which has a larger cross-section than pure gas (e.g., \citealt{czerny2015,baskin2017}).  Our results are at least qualitatively consistent with this model, with some important caveats.  For a wind that is viewed primarily in emission -- by force, primarily on the far side of the central engine -- {\em not} to appear significantly redshifted, its velocity must be low.  This effectively requires that the Fe~K line flux must originate close to the base of the wind, before it is accelerated (this is broadly consistent with a dusty origin for the BLR, \citealt{czerny2015}).

We note that an equally distinctive lag is not identified via JAVELIN when the Fe~K line flux alone--- via, e.g., a Gaussian model component---is used, rather than the entire reflection component.  Peaks within the lag spectrum are broad, and indistinct from other potential lags.  This is likely due to the limited sensitivity of our data.  It may also be due to related difficulties in isolating the line from the continuum, which includes both direct power-law flux and reflected flux.  In the limit of data with modest sensitivity, it is likely more pragmatic to rely on models that self-consistently account for the Fe~K line and reflected continuum.  Searches for lags between the direct power-law flux and continuum bands that exclude the Fe K region also do not show distinctive lags.

In the near future, observations of NGC 4388 with XRISM (\citealt{tashiro2018})  can effectively test and extend all of our results.  The calorimeter spectrometer aboard XRISM is expected to deliver a resolution of just 5~eV ($E/\Delta E \simeq 1300$ at the Fe~K$\alpha$ line),  coupled with a collecting area 10 times greater than the Chandra HEG.  Moderately deep observations with XRISM will reveal the details of the dynamical content of the Fe~K line in NGC 4388, and may reveal a complex of lines associated with distinct geometries.  At such high sensitivity, any Compton shoulder on the Fe~K line will be readily detected, enabling a clear measurement of the gas column density.  As the mission moves beyond initial observations of numerous targets, it may be possible to monitor NGC 4388 with XRISM to explore the line properties while also testing the tentative lag that we have found.

The maser emission in NGC 4388 previously enabled a precise measurement of its black hole mass (\citealt{kuo2011}).  However, the situation is not so fortunate in many other Seyfert-2 AGN, wherein there is no maser emission and the optical BLR is not visible.  Our results suggest that black hole masses in these systems could plausibly be measured by adapting optical BLR reverberation mapping techniques, replacing the optical continuum and, e.g., H$\beta$ line with the X-ray continuum reflection spectrum and narrow Fe~K line.  We expect that such efforts will need to be limited to Compton-thin AGN, as the line emission in Compton-thick AGN may be dominated by large-scale outflows and shaped by related motions (e.g., \citealt{kallman2014}).

\acknowledgements
This paper employs a list of Chandra datasets, obtained by the Chandra X-ray Observatory, contained in the Chandra Data Collection (CDC). \href{https://doi.org/10.25574/cdc.208}{doi:10.25574/cdc.208}.

We thank the anonymous referee for comments that improved this manuscript.  We acknowledge helpful conversations with David Bogensberger and Xin Xiang.

\bibliography{paper}
\bibliographystyle{aasjournal}

\end{document}